\documentclass[11pt,a4paper]{article}
\usepackage{jinstpub}

\newcommand{\TP}{\ensuremath{\mathrm{TP}}}
\newcommand{\TN}{\ensuremath{\mathrm{TN}}}
\newcommand{\FP}{\ensuremath{\mathrm{FP}}}
\newcommand{\FN}{\ensuremath{\mathrm{FN}}}

\newcommand{\mindrt}{\ensuremath{\Delta R(J,t)}}
\newcommand{\mindrw}{\ensuremath{\Delta R(J,W)}}

\newcommand{\ttbar}{\ensuremath{{t\bar{t}}}}

\newcommand{\pt}{\ensuremath{{p_\mathrm{T}}}}
\newcommand{\ptJ}{\ensuremath{{p_\mathrm{T}^J}}}

\newcommand{\Delphes}{\textsc{Delphes}\xspace}
\newcommand{\Pythia}{\textsc{Pythia}\xspace}

\newcommand{\MadGraph}{\textsc{MadGraph5}\xspace}

\title{Application of Machine Learning Based Top Quark and $W$ Jet Tagging to Hadronic Four-Top Final States Induced by SM and BSM Processes}

\usepackage{xspace}
\usepackage{comment}
\usepackage{subcaption}
\usepackage{graphicx}

\newcommand{\jt}[1]{\textcolor{black}{{#1}}}
\newcommand{\rv}[1]{\textcolor{black}{{#1}}}

\author[a]{Jiří~Kvita}
\author[a]{Petr~Baroň}
\author[b]{Monika~Machalová}
\author[a]{Radek~Přívara}
\author[b]{Rostislav~Vodák}
\author[b]{Jan~Tomeček}
\affiliation[a]{Joint Laboratory of Optics of Palacký University Olomouc and Institute of Physics of Czech Academy of Sciences, Czech Republic}
\affiliation[b]{Department of Mathematical Analysis and Applications of Mathematics, of Palacký University Olomouc, Czech Republic}

\emailAdd{petr.baron@upol.cz}
\emailAdd{rostislav.vodak@upol.cz}

\keywords{Machine Learning, Jet Structure, High Energy Physics}

\abstract{

  We apply both cut-based and machine learning techniques using the same inputs to the challenge of hadronic jet substructure recognition, utilizing classical subjettiness variables within the \Delphes{} parameterized detector simulation framework. 
  We focus on jets generated in simulated proton-proton collisions, identifying those consistent with the decay signatures of top quarks or $W$ bosons. Such jets are employed in four-top quark events in fully hadronic final states stemming from both the Standard Model as well as from a new physics process of a hypothetical scalar resonance $y_0$ decaying into a pair of top quarks. We reconstruct the resonance invariant mass and compare it properties over the falling background using the two tagging approaches, with implications to LHC searches.

}

\begin{document}
\maketitle

\section{Introduction}
Machine learning (ML) techniques are getting growing application in many research areas such as objects and events classification in high energy physics (HEP). 
The structure of this paper is as follows.
We first present a pedagogical overview of the application of selected basic ML techniques to the recognition of a substructure of hadronic final states (jets) and their tagging based on their possible origin in current HEP experiments using simulated events and a parameterized detector simulation. 

We present the samples, their jet composition, and results of per-jet tagging. We describe the truth labelling, sets used in training and testing as well as optimization of undersampling methods needed to train the ML algorithms.
We then check the tagging efficiencies and apply the trained taggers to dedicated samples with a clear signature of a jet mass peak. We compare to standard cut-based methods with the same inputs used and compare the physics performance as well as correlations between the tagging methods.

Finally we apply both cut-based and ML-based tagging methods to jet classification in four top quark final states, evaluating their performance on the reconstruction of a resonance from an extension of the Standard Model decaying to a pair of top quarks in the complex full-hadronic final state, with implications to current searches for hadronically decaying four-top final states in proton-proton collisions.

\section{Hadronic final states in high energy physics collisions}

Jets as hadronic final states are an inevitable consequence of the quantum chromodynamics (QCD)~\cite{Gross:2022hyw}, 
the force between strongly interacting matter constituents of quarks and gluons. In hadron collisions, jets are important final states and signatures of objects of high transverse momentum. 

In cases of large jet transverse momenta, i.e. with a large Lorentz boost in the plane perpendicular to the proton beam, decay products of hadronically decaying $W$ bosons or top quarks are collimated so that they form one large boosted jet in the detector. 
Large jets high transverse momentum phase space region is of special interest due to its gradual appearance with growing luminosity of current accelerators like the LHC, offering windows to test QCD in new kinematic regions, but also due to the possible existence of heavy new physics resonances decaying to top quark pairs, leading to highly boosted top quarks or $W$ bosons.

The varying jets substructure of hadronic jets of different origin is a key feature exploited in tagging of jets as coming from the hadronically decaying $W$ or $Z$ bosons, the top quark ($t$), or even the Higgs boson ($H$), with their physical masses being actually measured as $m_W \doteq 80.37\,\mathrm{GeV}$, 
$m_Z \doteq 91.19 \,\mathrm{GeV}$, $m_t \doteq 172.69\,\mathrm{GeV}$ and $m_H \doteq 125.25\,\mathrm{GeV}$~\cite{Workman:2022ynf}.
Many jets are of a non-resonant origin, giving a rise of a continuum in the jet mass spectrum ($m_J$) but of much larger yield then a weak signal.

Hadronic jets appear as signatures of many new physics final states as well. In this paper we shall explore the four-top quark final state ($\ttbar{}\ttbar{}$, or $4t$) produced both within the Standard Model (SM) as well as via a benchmark process beyond the SM (BSM). The four-top quark production in fully hadronic final states is receiving more and more attention also from the theoretical point of view~\cite{Jezo:2021smh}.

\section{Simulations}

\subsection{Samples generation}

Both Standard Model and Beyond-the-Standard-Model samples were simulated for this study as a source of events with hadronic final states. Using the \MadGraph{} version {\tt 2.6.4} simulation toolkit~\cite{Alwall:2014hca}, proton-proton collision events at $\sqrt{s} = 14$ TeV were generated for the SM process $pp \rightarrow \ttbar{}$ in the all-hadronic $\ttbar$ decay channel at next-to-leading order (NLO) in QCD in production, 
using the MLM matching~\cite{Hoeche:2005vzu,Mangano:2002ea}, \emph{i.e.}~with additional processes with extra light-flavoured jets produced in the matrix element, matched and resolved for the phase-space overlap of jets generated by the parton shower using \MadGraph{} defaults settings. 
The parton shower and hadronization were simulated using \Pythia{}8~\cite{Sjostrand:2014zea}.

As a train BSM model, the resonant $s$-channel \ttbar{} production via an additional narrow-width (sub-GeV) vector boson $Z'$ as $pp \rightarrow Z' \rightarrow \ttbar{}$ using the model \cite{FeynModelZprime,Christensen:2008py,Wells:2008xg}
were generated, to provide a sample of top quarks with large transverse momenta, enhancing the boosted regime.

As a representative model of a BSM process for testing, the production of a scalar resonance decaying to a pair of top quarks $y_0 \rightarrow \ttbar{}$ was adopted~\cite{Christensen:2008py} at the leading-order (LO) 
in the $\ttbar$ production with the gluon-gluon fusion loop 
(more details in~\cite{Mattelaer:2015haa,Backovic:2015soa,Neubert:2015fka,Das:2016pbk,Kraml:2017atm,Albert:2017onk,Arina:2017sng,Afik:2018rxl}), 
with inclusive $\ttbar{}$ decays, selecting the all-hadronic channel later in the analysis. 

\subsection{Parameterized detector simulation}
Using the \Delphes{} (version {\tt 3.4.1}) detector simulation \cite{deFavereau:2013fsa} with the ATLAS card, jets with distance parameters of $R=1.0$ (dubbed as large-$R$ jets) were reconstructed using the anti-$k_t$ algorithm using the FastJet package~\cite{Cacciari:2011ma} at both particle and detector levels.

The trimming jet algorithm \cite{Krohn:2009th} as part of the \Delphes{} package was used to obtain jets with removed soft components, using the parameter of $R_\mathrm{trim} = 0.2$ and modified \pt{} fraction parameter $f^{\pt}_\mathrm{trim} = 0.03$ (originally 0.05). The trimming algorithm was chosen over the standard non-groomed jets, soft-dropped~\cite{Larkoski:2014wba} and pruned jets~\cite{Ellis:2009me}, with parameters varied, in terms of the narrowness of the mass peaks.

\subsection{Objects of interest}
The interest is the identification of large-$R$ hadronic jets coming from the hadronic decays of top quarks and $W$ bosons. In the na\"{i}ve picture of the hadronic decays of $W \rightarrow q\bar{q}'$ and $t \rightarrow W b \rightarrow b q\bar{q}'$, these manifest themselves as three and two prong decays, respectively. Different jet substructure is thus expected for such $t$ and $W$ jets.

\section{Cut-based top and $W$ tagging}
As input variables to both cut-based as well as ML-based tagging we utilize simple yet powerful ``classical'' variable called $n$-subjettiness~\cite{Thaler_2011}, $\tau_N$, which is related to the consistency of a jet with the 
hypothesis of containing $N$ subjets. These variables are combined into ratios $\tau_{32}$ and $\tau_{21}$, defined as $\tau_{ij} \equiv \frac{\tau_i}{\tau_j}$.
We thus compare cut-based and ML-based methods using the same inputs.

In order to identify jets coming from the hadronic decays of the $W$ boson or a top quark by a simple cut-based algorithm, large-$R$ jets were tagged as
\begin{itemize}
\item $W$-jets if $ 0.10 < \tau_{21} < 0.60 \, \land \,   0.50 < \tau_{32} < 0.85 \, \land \,  m_J \in [60, 110] \,\mathrm{GeV}$;
\item top-jets if $ 0.30 < \tau_{21} < 0.70 \, \land \,   0.30 < \tau_{32} < 0.80 \, \land \,  m_J \in [138, 208] \,\mathrm{GeV}$.
\end{itemize}

\begin{figure} 
\centering
\begin{subfigure}{0.48\textwidth}
    \includegraphics[width=\linewidth]{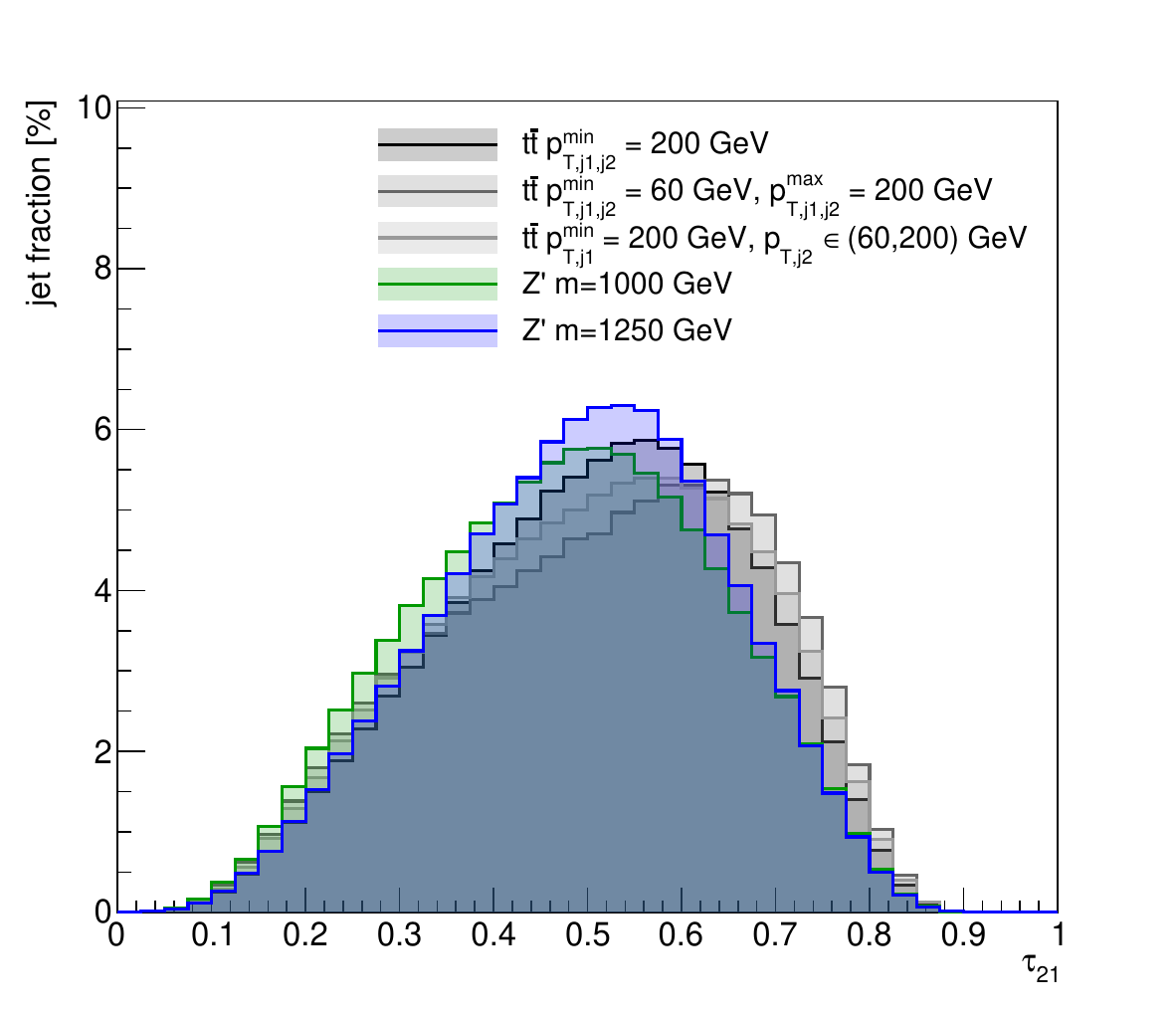}
    \caption{} \label{fig:1a}
  \end{subfigure}%
  \hspace*{\fill}   
  \begin{subfigure}{0.48\textwidth}
    \includegraphics[width=\linewidth]{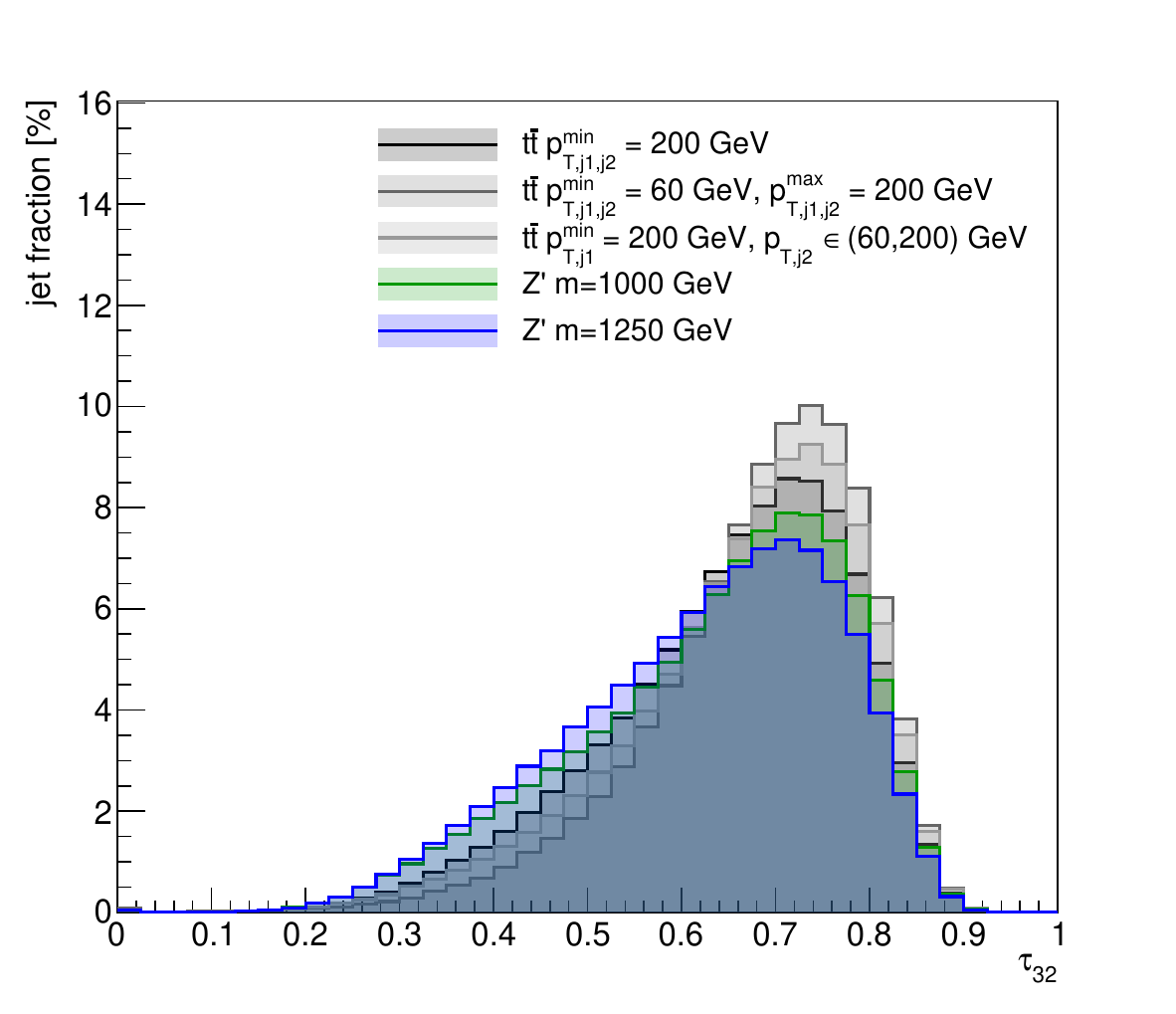}
    \caption{} \label{fig:1b}
  \end{subfigure}%
  \\ 
  \begin{subfigure}{0.48\textwidth}
    \includegraphics[width=\linewidth]{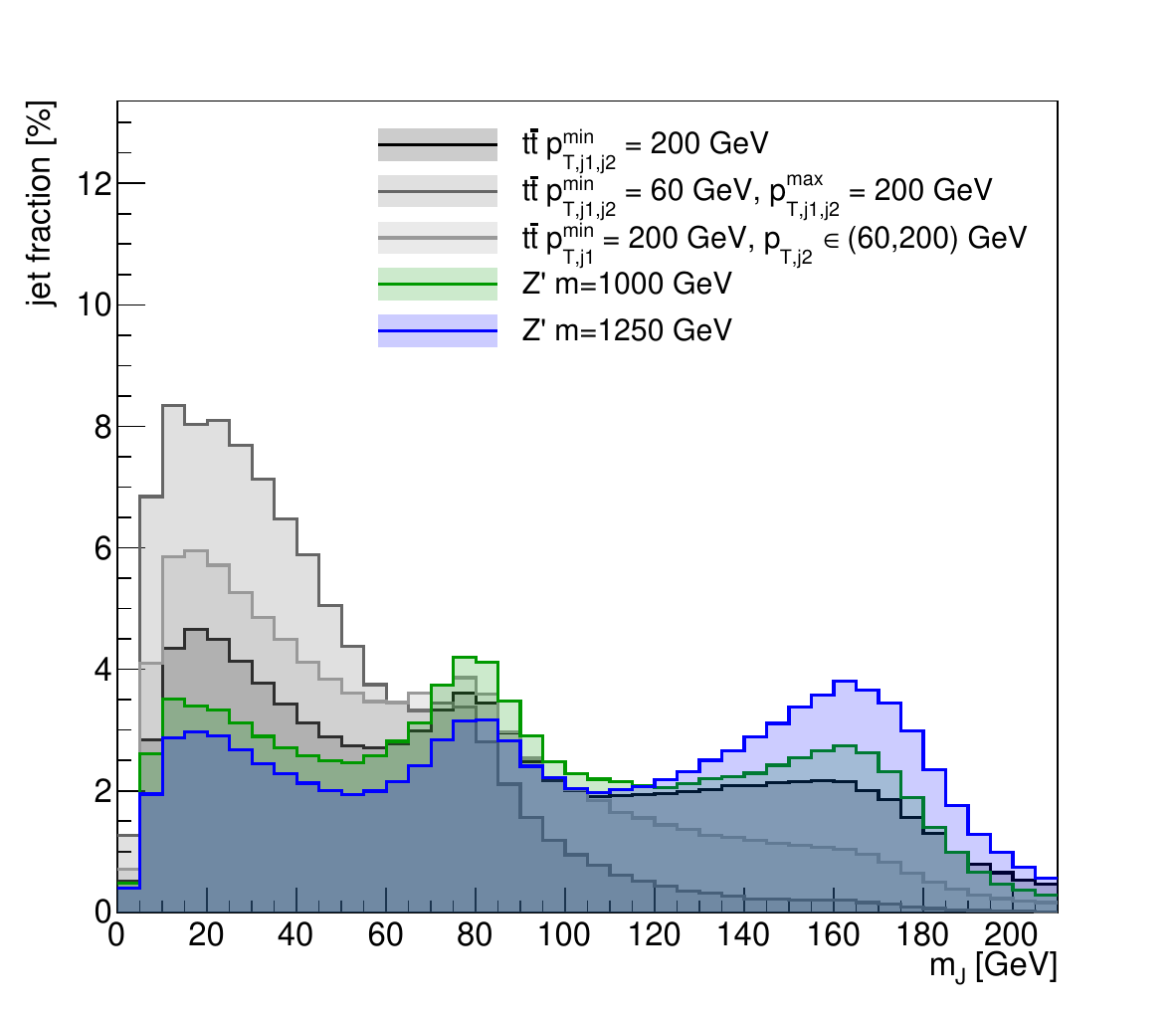}
    \caption{} \label{fig:1c}
  \end{subfigure}

\caption{Shapes of the $\tau_{21}$, $\tau_{32}$ subjettiness variables (top) and the large-$R$ jet mass (bottom) in the five samples used in training and testing of the tagging algorithms.}
\label{fig:mass_taus_samples}
\end{figure}

Shapes of the variables used as input to the ML classifier are shown in Figure~\ref{fig:mass_taus_samples} for the individual samples.
One can observe the enhancement in the $Z'$ samples at the place of the expected top quark mass peak, the larger the higher the mass of the $Z'$ particle, while the lower mass $Z'$ sample provides enhanced region at the $W$ boson mass. The various \ttbar{} samples exhibit a large continuum of masses, with non-resonant bulk contribution below 60~GeV of different sizes due to different jet \pt{} kinematics cut for the samples.

\section{ML-based top and $W$ tagging}

Three samples corresponding to the SM \ttbar{} production were generated, with different cuts at the generator level on the transverse momentum of the jets, in order to cover regions with various fractions of $t$, $W$ as well as non-resonant (light) jets. These have been used as both training and testing data sets. 

The two $Z'$ samples with the $Z'$ masses of $1000$ and $1250\,$GeV provide a $\ttbar{}$ sample with enhanced boosted top quarks, thus leading to events with enhanced fractions of $t$ and $W$ jets.

\noindent
Variables defined and used for each jet in the classification are as follows
\begin{itemize}
    \item Jet transverse momentum $\ptJ$ and jet four-vector invariant mass $m_J$.
    \item $\eta$  and $\phi$ of the jet.
    \item Jet substructure variables $\tau_{32}$ and $\tau_{21}$.
\end{itemize}
\noindent

Variables used to define the truth labelling are as follows
\begin{itemize}
    \item \mindrw{}, the minimal angular separation of the jet to the nearest $W$ \footnote{The angular distance between two objects is defines as $\Delta R \equiv \sqrt{ (\Delta\phi)^2 + (\Delta\eta)^2}$ where the pseudorapidity $\eta \equiv -\ln\tan\frac{\theta}{2} $ is related to the standard azimuthal angle $\theta$ of the spherical coordinates, where the beam axis coincides with the $z$ axis, and $\phi$ is the polar angle in the $xy$ plane.}; 
    \item \mindrt{}, the minimal angular separation of the jet to the nearest top parton. 
\end{itemize}

The true type jets labels are then based on the following criteria
\begin{enumerate}
\item truth $t$-jets:  $\mindrt{}<0.1 \land 138\,\mathrm{GeV}\leq m_J \leq 208\,\mathrm{GeV}$;
\item truth $W$-jets; $\mindrw{}<0.1 \land 60\,\mathrm{GeV}\leq m_J \leq 100\,\mathrm{GeV}$;
\item truth light jets: otherwise.
\end{enumerate}
\rv{For training and testing, the variables \mindrt{} and \mindrw{} are excluded from the processes as they are not available in real data at the detector level.}

\begin{table}[h] 
\centering
\begin{tabular}{c|l|c|c}
Sample ID & Sample definition & Number of jets & Events\\
\hline
1 & \ttbar{} $p_\mathrm{T,j1,j2}^\mathrm{min} = 200\,$GeV & 797k & 317k\\
2 & \ttbar{} $p_\mathrm{T,j1,j2}^\mathrm{min} = 60\,$GeV, $p_\mathrm{T,j1,j2}^\mathrm{max} = 200\,$GeV  & 447k & 236k\\
3 & \ttbar{} $p_\mathrm{T,j1}^\mathrm{min} = 200\,$GeV, $p_\mathrm{T,j2} \in [60,200]\,$GeV & 782k & 325k\\
4 & $Z'$, $m = 1000\,$GeV & 450k & 274k\\
5 & $Z'$, $m = 1250\,$GeV & 389k & 213k\\
\end{tabular}
\caption{The \ttbar{} samples definition for training and testing and the number of events in each dataset.}
\label{tab:sample_names_stats}
\end{table}

\begin{figure}[h] 
\centering
\includegraphics[width=0.65\textwidth]{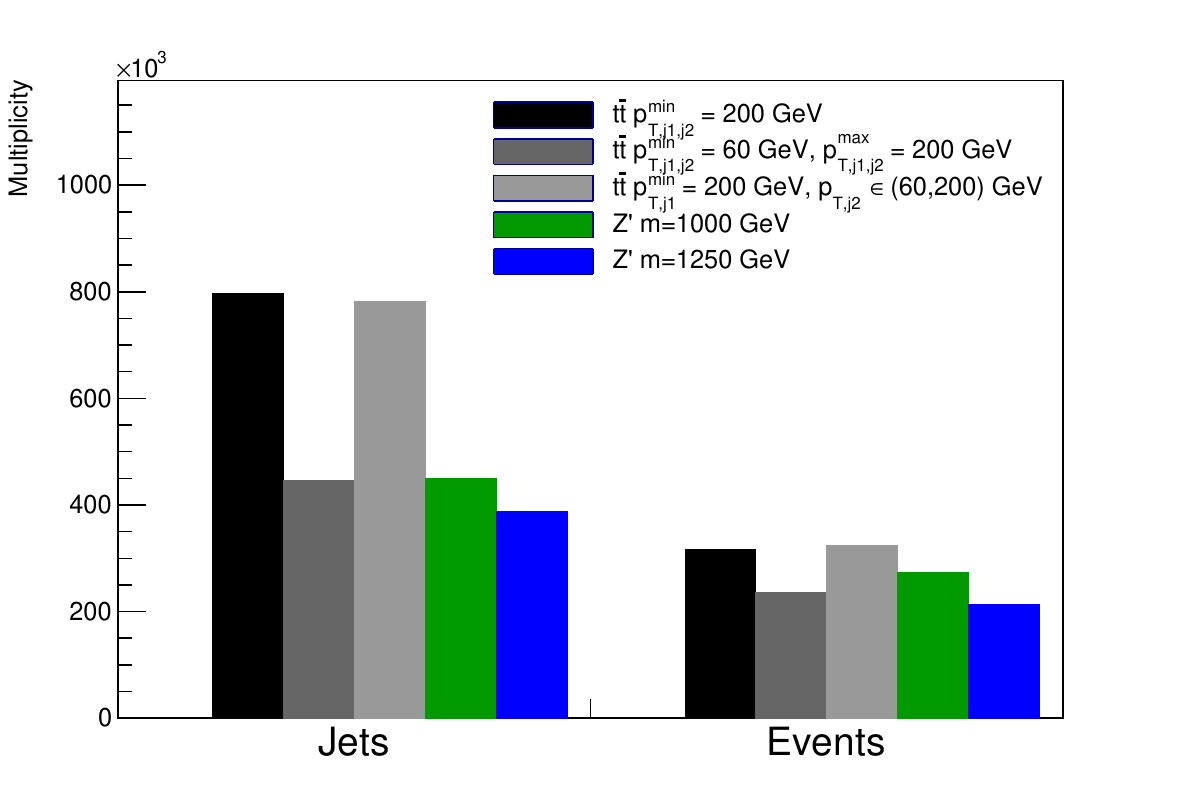}
\caption{The number of events and events in each dataset used for training and testing.}
\label{fig:sample_names_stats}
\end{figure}

\begin{figure}[h]
\centering
\includegraphics[width=0.65\textwidth]{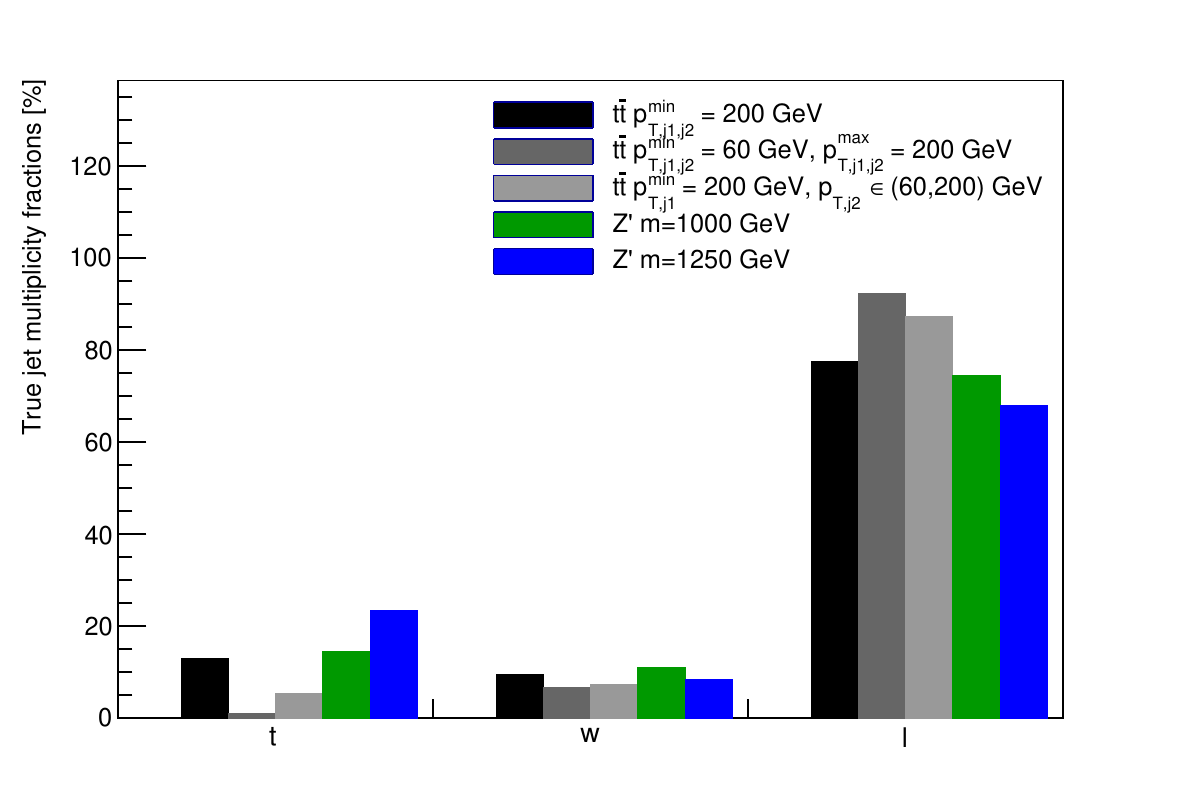}
\caption{Fractions of true labels of jets in samples (in \%).}
\label{table:stats1}
\end{figure}
The datasets samples definitions and structure is in Table~\ref{tab:sample_names_stats}, with indicated number of events and jets in the samples. The same information is also displayed graphically in~Figure~\ref{fig:sample_names_stats}.

The jet labels of $t$, $W$ and light ($l$) jets correspond to the definition above. In Figure~\ref{table:stats1} we summarize jets proportions (multiplicities) in the data sets.
\begin{figure}[h]
\centering
\includegraphics[width=0.65\textwidth]{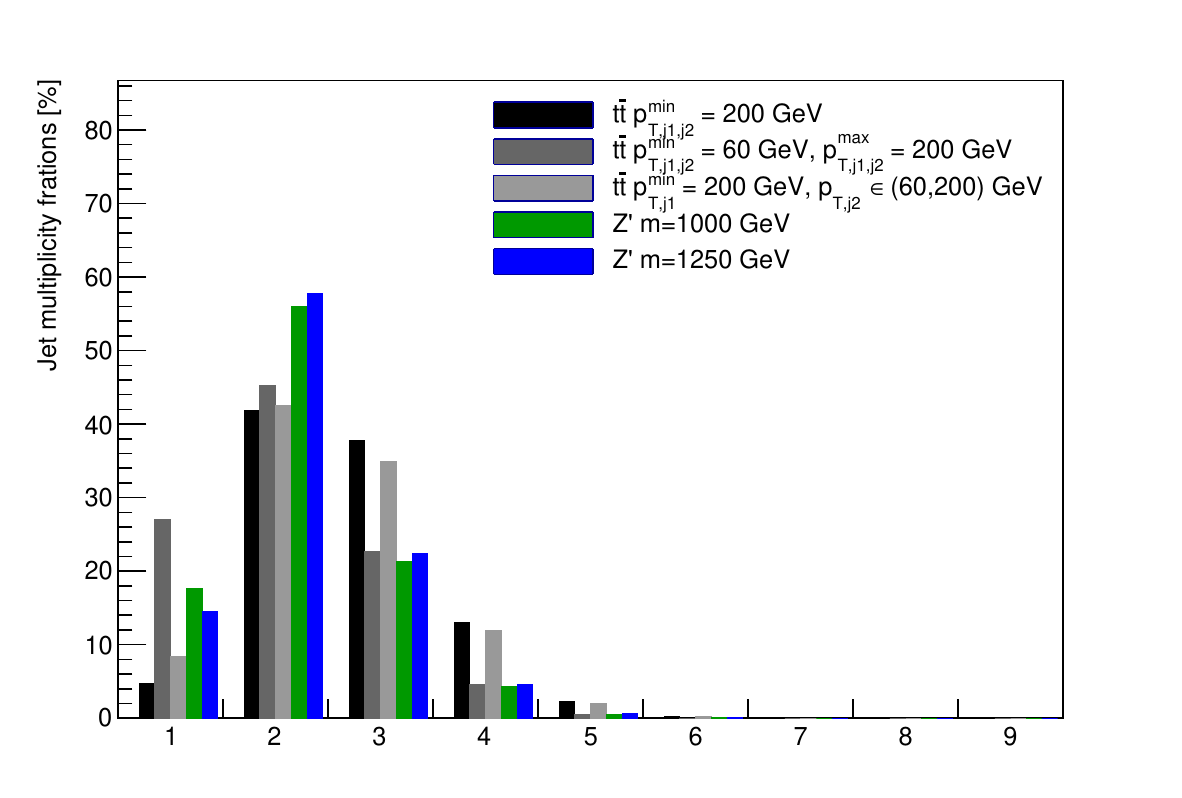}
\caption{Proportions of the events with various number of jets (in \%).}
\label{tab:prop_jets}
\end{figure}


\subsection{Structures of data sets}

\rv{The data sets in Table~\ref{tab:sample_names_stats} were used to create two final data sets}. \rv{The first one is the unification of the $Z'$ sets (IDs 4 and 5) and the second one is the unification of the $\ttbar{}$ sets (IDs 1--3). 
From the above criteria, it is clear that to identify particular jets one can restrict one's attention to the jets with mass in intervals $[60,100]$ GeV and $[138,208]$ GeV. Otherwise, the jets are light jets by definition. 
The ratios of respective jets are summarized in the following tables}
\jt{
\begin{itemize}
\item Samples used for the $t$-jets identification:
\begin{center}
\begin{tabular}{|c|c|c|}
\hline
{\bf Data set} & {\bf $t$-jets} & {\bf light-jets}\\
\hline
$Z'$ $t$-set      & 78\%& 22\%\\
$\ttbar{}$ $t$-set & 62\% & 38\%\\
\hline
\end{tabular}
\end{center}
\item Samples used for $W$-jets identification:
\begin{center}
\begin{tabular}{|c|c|c|}
\hline
{\bf Data set} & {\bf $W$-jets} & {\bf light-jets}\\
\hline
$Z'$ $W$-set  & 42\% & 58\%\\
$\ttbar{}$ $W$-set & 35\% & 65\%\\
\hline
\end{tabular}
\end{center}
\end{itemize}
}

\subsection{Preprocessing}
\rv{For preprocessing we use {\tt scikit-learn} library (see Section 5.5) with respective classes. The data sets from the previous section were decomposed into the training and the test sets using the class 
\texttt{StratifiedShuffleSplit()} which ensures that the training and the test sets have the same ratios of $t$-jets, $W$-jets and light-jets as the original sets. The training sets contain $80\%$ and the test sets $20\%$ of data from the original sets. We further use the class \texttt{StandardScaler()} which scales all features according to the relation} 
\rv{$$z = \frac{x - \mu}{\sigma}\,,$$}
\rv{where $\mu$ is the mean of the training samples and $\sigma$  is the standard deviation of the training samples. 
The respective transformations based on the scalings were then applied to the test sets. The reason for the scaling was 
that we also use neural networks for a tagging which do not work very well in the case when the features have very 
different scales.} 

\rv{It is evident from the table above is that the ratio between $t$-jets and light-jets is very distorted in the direction of $t$-jets. As a result, machine learning methods tend to ignore the minor class and label all instances according to the major class. There are several ways how to treat the case. Due to the sufficient amount of data, we settled for the undersampling applied to the training sets, which uses various techniques to remove data from the major class. Its advantage is that it 
does not add any artificial information to data compared with oversampling. We tested the following techniques of undersampling:
\begin{itemize}
\item {\it Random undersampling} under-samples the majority class by randomly picking samples with or without 
replacement.
\item {\it Cluster centroids} \cite{Yen:2009} undersamples by generating centroids based on clustering methods.
\item {\it Near miss} \cite{Mani:2003} is an algorithm based on NearMiss methods, selecting samples from the majority 
class for which the average distance of the k nearest samples of the minority class is the smallest.
\item {\it Repeated edited nearest neighbor (ENN) method} \cite{Tomek:1976} is a method is based on the ENN method that works by finding the $k$-th nearest neighbor of each observation first, then checking whether the majority class from the observation’s $k$-th nearest neighbor is the same as the observation’s class or not.
\end{itemize}
}

\subsection{Methodology}
\rv{
In the process of evaluation, we calculate the following four basic metrics
\begin{eqnarray}
\mathrm{Accuracy} &\equiv& \frac{\TP + \TN}{\TP + \TN + \FP + \FN}
\label{1} \\
\mathrm{Precision} &\equiv& \frac{\TP}{\TP + \FP}
\label{2} \\
\mathrm{Recall} &\equiv& \frac{\TP}{\TP + \FN} \equiv \mathrm{True\ positive\ rate} \equiv \epsilon_\mathrm{tag} \label{3} 
\\
\mathrm{False\ positive\ rate} &\equiv& \frac{\FP}{\FP + \TN} \equiv \epsilon_\mathrm{mistag} \label{4} \,,
\end{eqnarray}
where $\TP$ stands for {\bf true positive\/}, $\TN$ for {\bf true negative\/},
 $\FP$ for {\bf false positive\/} and $\FN$ for {\bf false negative\/}. }
 
 \rv{For the predictions we use the two machine learning (ML) models that rank among the best, namely
\begin{itemize}
\item {\it Gradient boosting classifier} (GBC) is one of the two most 
 used types of \emph{ensemble methods}, which are methods combining multiple simple predictors (here decision trees)  to create a more powerful model. The method does not work with weights but it tries to fit the predictor to the  \emph{residual errors} made by the previous predictor. The new prediction is made by adding up all the predictors'  predictions \cite{MullerGuido}.
\item {\it Multi-layer Perceptron classifier} (MLP) is a classifier based on artificial neural networks. 
\end{itemize}
}

\rv{We use the grid search for both algorithms to tune their hyper-parameters. In the case of gradient-boosting classifier, we tune the following hyper-parameters:
\begin{itemize}
\item the number of estimators;
\item the function to measure the quality of a split;
\item maximum depth of the individual regression estimators;
\item the number of features to consider when looking for the best split.
\end{itemize}
}
\rv{In the case of MLP, we tune the following hyper-parameters
\begin{itemize}
\item  the number of hidden layers;
\item activation functions;
\item learning rate;
\item strength of the $L^2$ regularization term.
\end{itemize}
We also applied early stopping and cross-validation to prevent overfitting. 
}

\subsection{Performance of ML algorithms}

\begin{figure}[!h]
    \centering
    \includegraphics[width=0.65\textwidth]{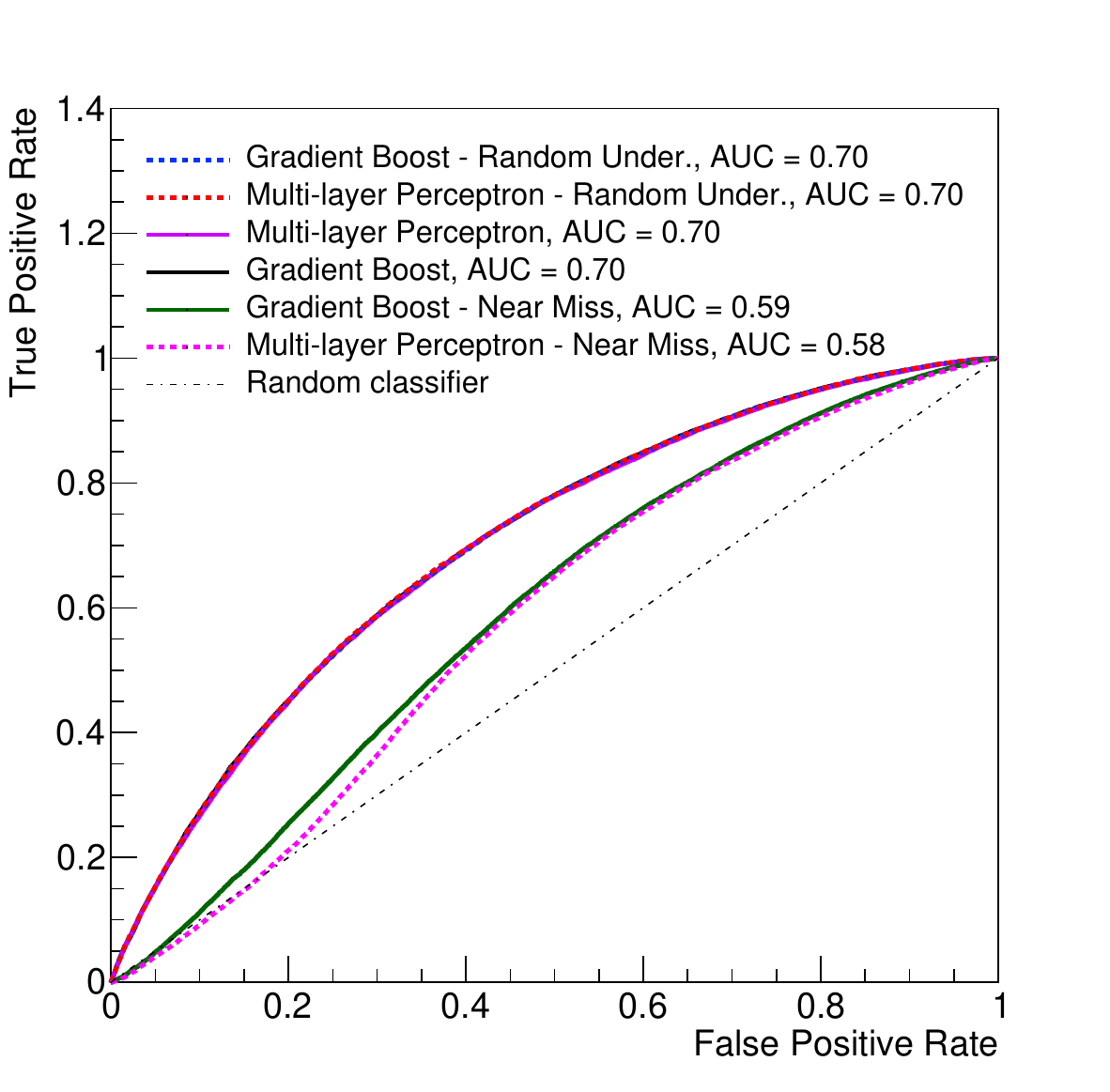}
    \caption{ ROC curves summarising the performance of $W$-tagging classifiers upon test samples.}
    \label{fig:rocs1}
\end{figure}

\rv{For training and testing the respective algorithms, we used different sets. The algorithms for the prediction of $t$-jets were trained, after applications of under-sampling methods,  on a part of the $Z'$ $t$-set and tested on the 
rest of $Z'$ $t$-set and $\ttbar{}$ $t$-set. Let us point out that the results below are for the whole training set and not for their parts given by the under-sampling methods. The algorithms for the prediction of $W$-jets were trained on a part of the $\ttbar$ $W$-set and tested on the rest of $\ttbar$ $W$-set and $Z'$ $W$-set.}
\rv{In the end, the GBC for the prediction of $W$-jets and GBC with random under-sampling for the prediction of $t$-jets was chosen with 
area under ROC curve (AUC) 0.70 for $W$-tagging and 0.67 for the $t$-tagging.} 
The performance of classifiers is shown via ROC curves derived based on test samples in Figure~\ref{fig:rocs1} for $W$-tagging and in Figure~\ref{fig:rocs2} for $t$-tagging. Detailed view on the performance of each of ML algorithm is given in~Appendix~\ref{app:perf}.

\begin{figure}[!h]
  \centering
  \includegraphics[width=0.65\textwidth]{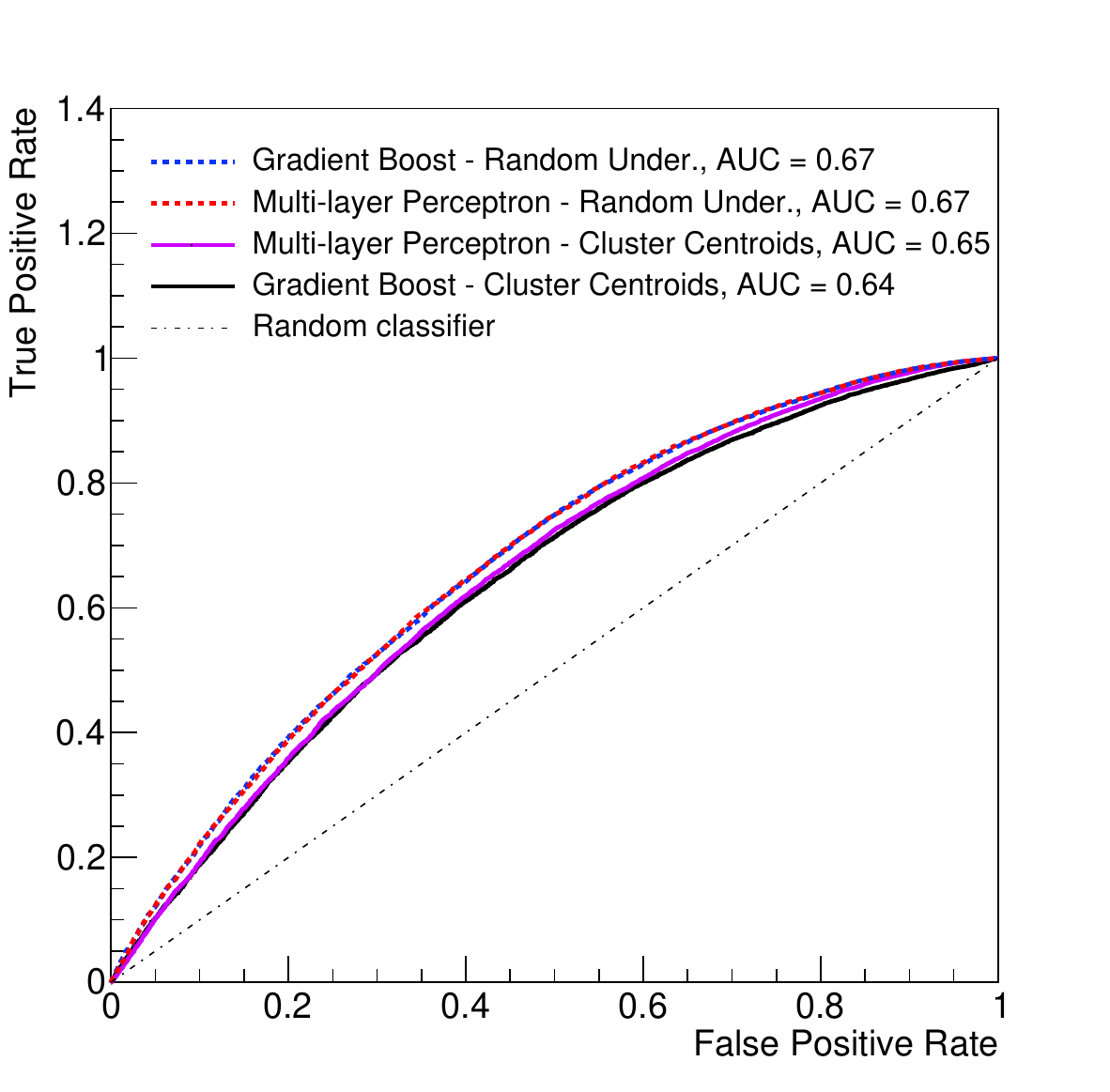}
  \caption{ ROC curves summarising the performance of $t$-tagging classifiers upon test samples.}
  \label{fig:rocs2}
\end{figure}

\subsection{Implementation and integration to a C++ based code}
\rv{The implementation of the model is carried out in C++ and the algorithms were trained in Python; in particular, we use the well-known open source ML library {\tt scikit-learn}~\cite{scikit-learn}. In detail, the Gradient Boosting Machines technique is implemented via the class {\tt sk\-learn\-.ensem\-ble\-.Gradient\-Boosting\-Classifier}. Multi-layer Perceptron classifier is implemented via the class {\tt sklearn.neural\_network.MLPClassifier}.} The user code in HEP is usually based on C++. The integration between these two languages is made by \texttt{pybind11}, which is a lightweight header library exposing C++ types in Python and vice versa, see \url{https://pybind11.readthedocs.io/en/stable/}. 

The Python code is contained in the module \texttt{in\_out.py}, where only the following three Python-functions are called from the C++ source code: 
\begin{itemize}
\item \texttt{load\_classifiers}: loads the trained classifiers (stored in enclosed \texttt{pickle}-files);
\item \texttt{evaluate}: the very prediction function; the input is a jet (six features: '$p_\mathrm{T}$', '$\eta$', '$\phi$', '$\tau_{32}$', '$\tau_{21}$', 'mass') and the output is its evaluation by the classifier (one of the values: 't', 'W', 'light'),
\item \texttt{evaluate\_mat}: the same functionality as \texttt{evaluate}, the input is a matrix of jets (better for predictions for more jets; it loads classifiers only once).
\end{itemize}

\section{Comparison of ML and cut-based tagging}

\subsection{Example of data points in $\tau_{21}$ and $\tau_{32}$ spaces}
\begin{figure}[!h]
    \centering
      \begin{subfigure}{0.48\textwidth}
        \includegraphics[width=\linewidth]{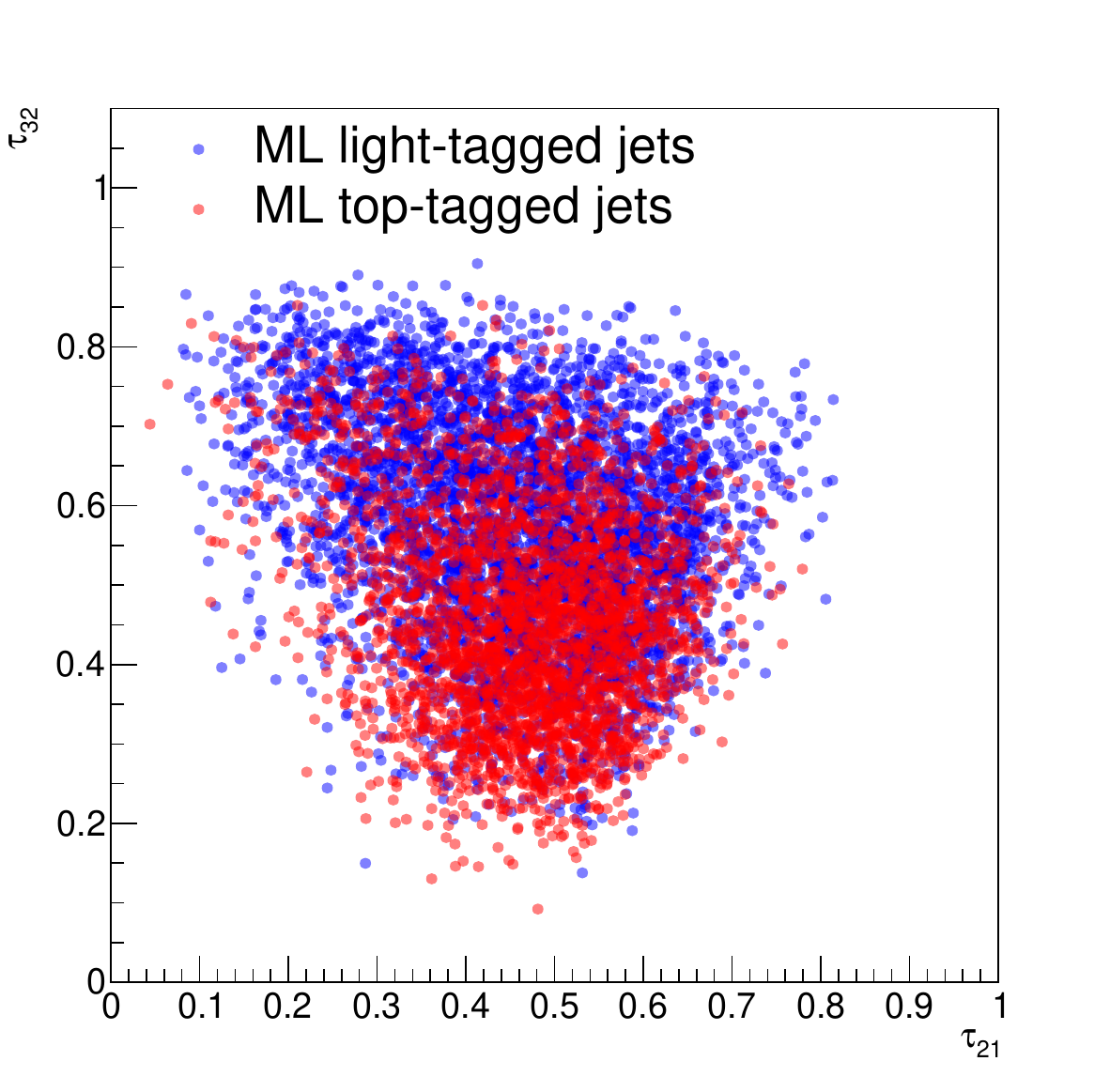}
        \caption{} \label{fig:1a_example}
      \end{subfigure}%
      \hspace*{\fill}   
      \begin{subfigure}{0.48\textwidth}
        \includegraphics[width=\linewidth]{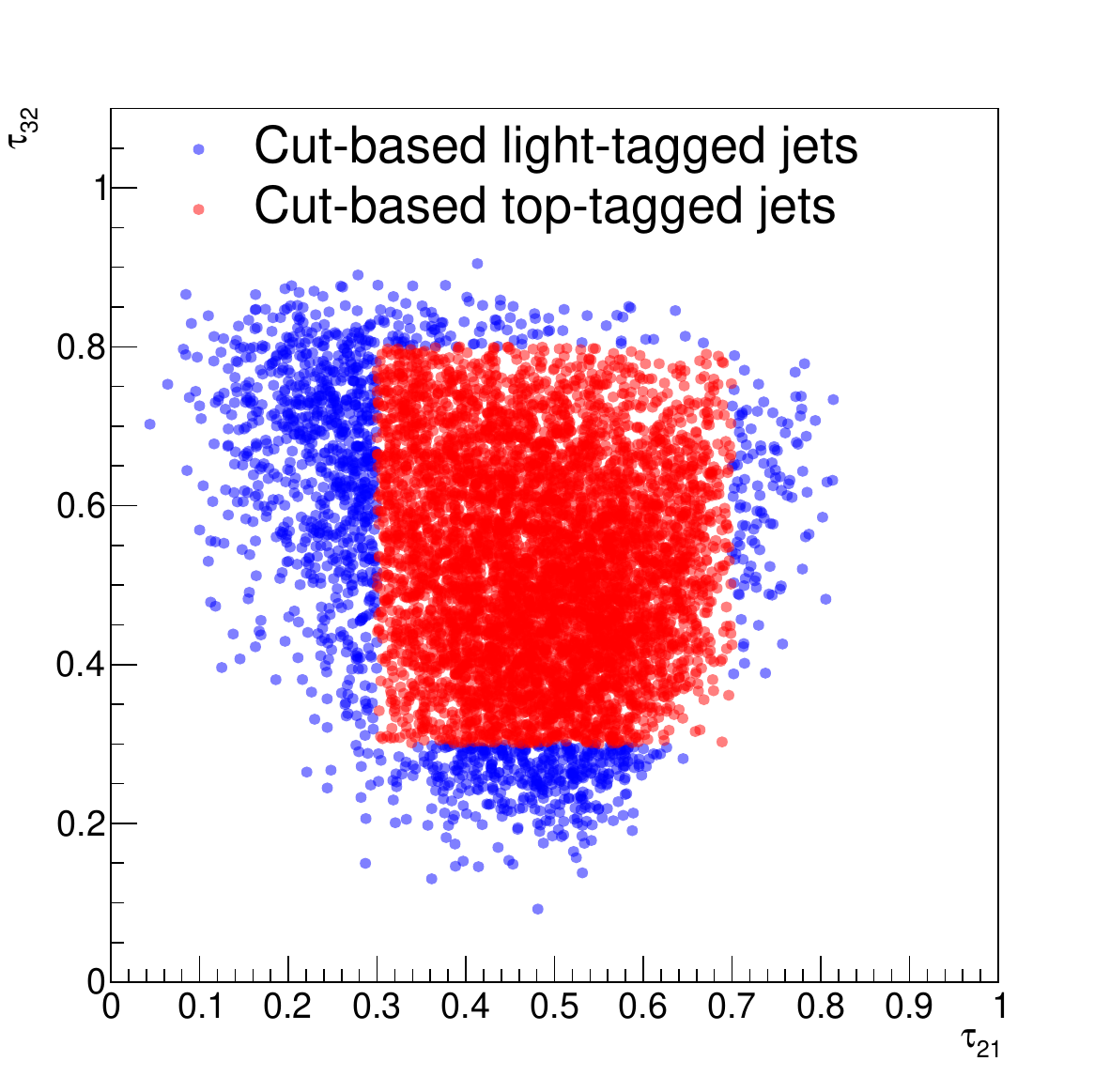}
        \caption{} \label{fig:1b_example}
      \end{subfigure}
      \\
      \begin{subfigure}{0.48\textwidth}
        \includegraphics[width=\linewidth]{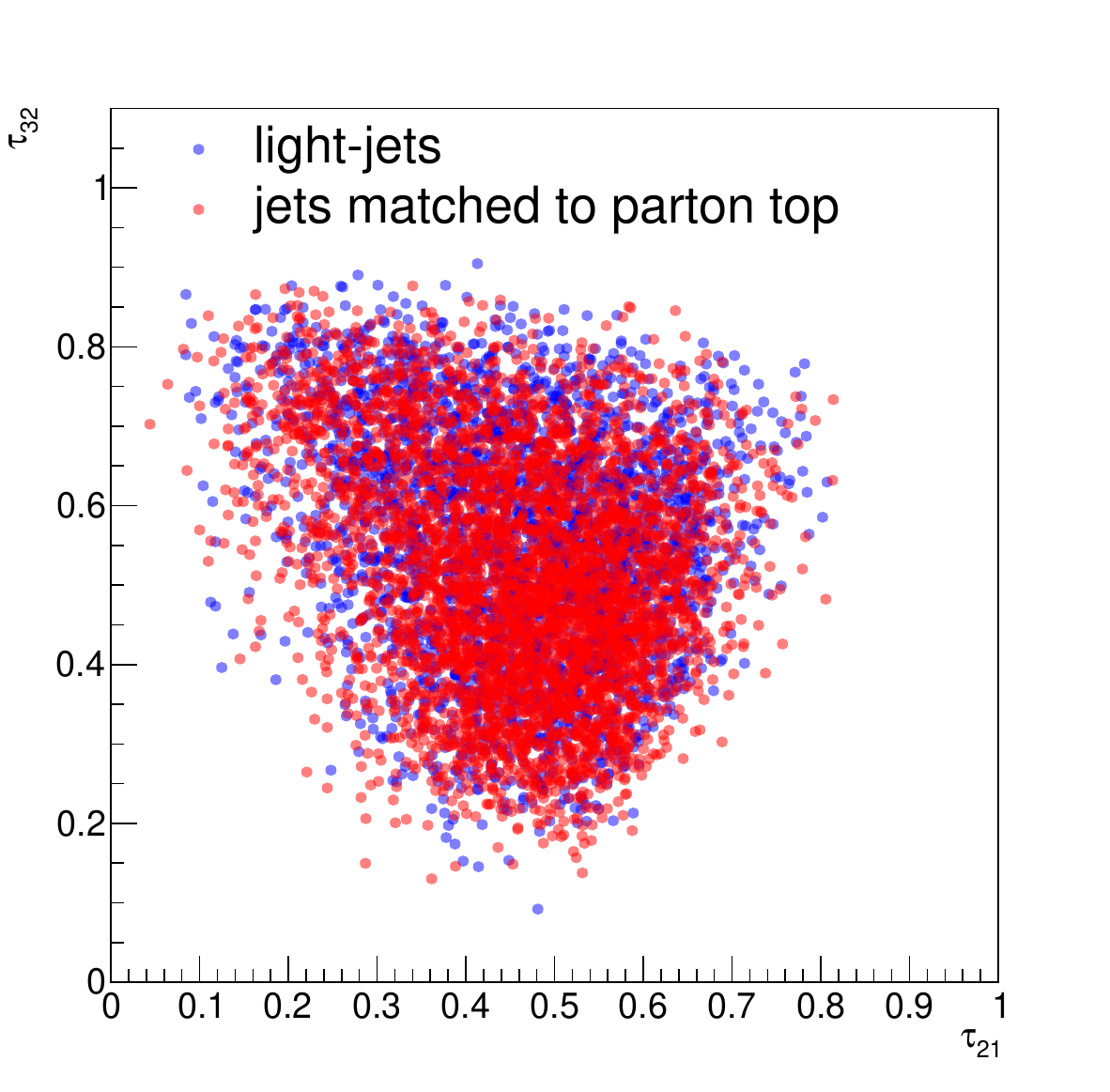}
        \caption{} \label{fig:1c_example}
      \end{subfigure}%
    \caption{Data points of SM $t\bar{t}$ subsample with background light jets - blue dotted points and signal top (a) tagged using ML, (b) tagged using cut-based method, and (c) matched to parton within $\Delta R < 0.1$.}
    \label{fig:points}
\end{figure}

The Figures~\ref{fig:1a_example} and \ref{fig:1b_example} present examples of 100k jets being classified as top jets (red) using ML-based and cut-based method, respectively. The blue points stands for the jets tagged as light jets. 
The cut-based method emerges as rectangle shape while ML-based approach is non-linear. 

The red points in the Figure~\ref{fig:1c_example} are the truth labels based on jet matching to top quark within $\Delta R < 0.1$. This $\tau_{21}$ versus $\tau_{32}$ projection indicates the challenge since no clearly visible pattern in separating $t$-jets (red) from light jets (blue) stands out. 

\subsection{Physics samples used}
Three more samples have been generated in order to test the tagging performance in more realistic applications.
First a jet sample coming purely from QCD interactions, thus exhibiting no resonance structure and ideal for checking the mistag rate was generated with varied thresholds on jets transverse momenta similarly to those of the \ttbar{} samples.
Then, a SM $4t$ sample was generated where all top quarks were forced to decay hadronically, leading to a sample with potentially a large number of true $W$ and $t$ jets.
An example BSM sample also with four top quarks in the final states but with one pair of top quarks coming from a decay of a scalar resonance of mass of $1500\,$GeV was also generated, in order to test the search for a resonant peak in the $\ttbar{}$ invariant mass spectrum within the $4t$ final state.

\subsection{Tagging efficiencies and mistagging rate}
In this section the real efficiencies $\epsilon_{\mathrm{real}}$ and mistagging rate (fake efficiencies) $\epsilon_{\mathrm{fake}}$ are plotted as a function of jet $p_T$ and mass. In each bin of the $p_{T}$ and mass distributions the particular bin content is given by 
Eq.\ref{eq:realeff} and Eq.\ref{eq:fakeeff}. As for the mistagging rate the QCD samples were used. 
\begin{equation}
  \epsilon_{\mathrm{real}} = \frac{\mathrm{N(tagged~\&~matched)}}{\mathrm{N(tagged~\&~matched)}+\mathrm{N(not-tagged~\&~matched)}}
  \label{eq:realeff}
\end{equation}

\begin{equation}
  \epsilon_{\mathrm{fake}} = \frac{\mathrm{N(tagged~\&~not-matched)}}{\mathrm{N(tagged~\&~not-matched)}+\mathrm{N(not-tagged~\&~not-matched)}}
  \label{eq:fakeeff}
\end{equation}

\begin{figure}[!h]
    \centering
    \begin{subfigure}{0.48\textwidth}
        \includegraphics[width=\linewidth]{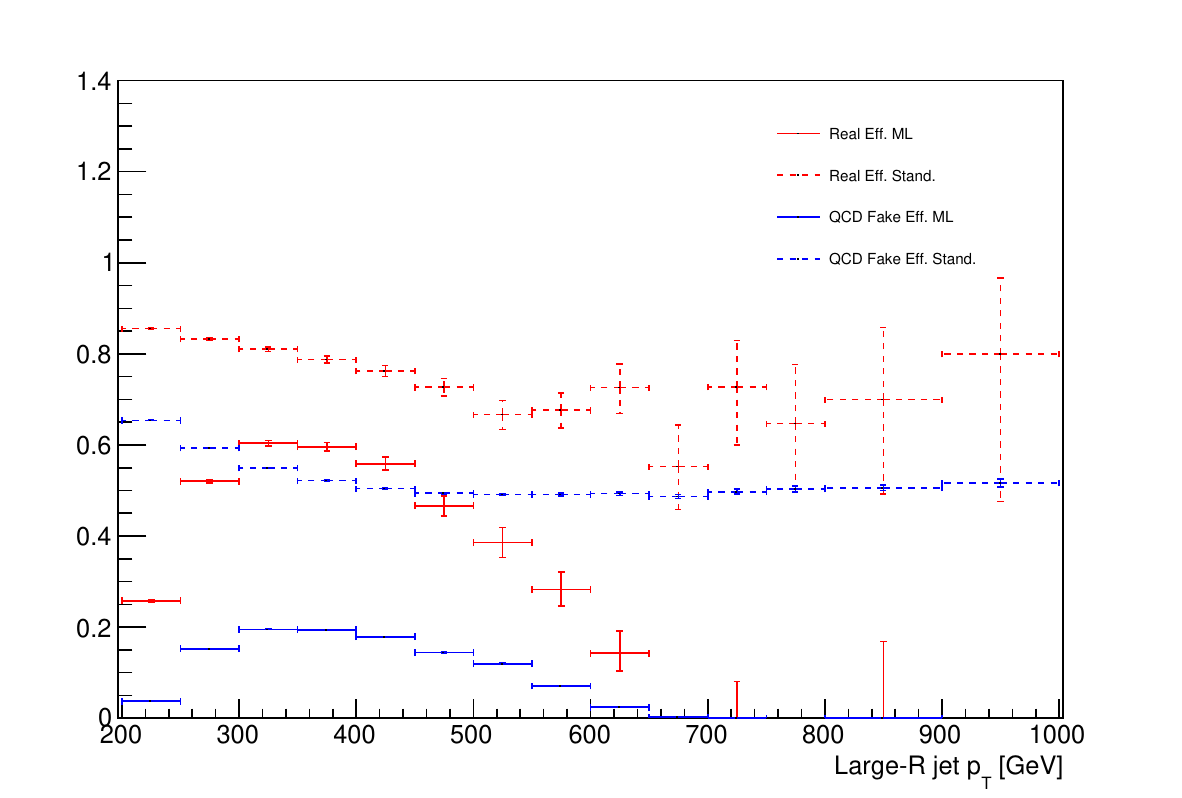}
        \caption{SM $t\bar{t}$.} \label{fig:1a_eff_smtt_w}
      \end{subfigure}%
      \hspace*{\fill}   
      \begin{subfigure}{0.48\textwidth}
        \includegraphics[width=\linewidth]{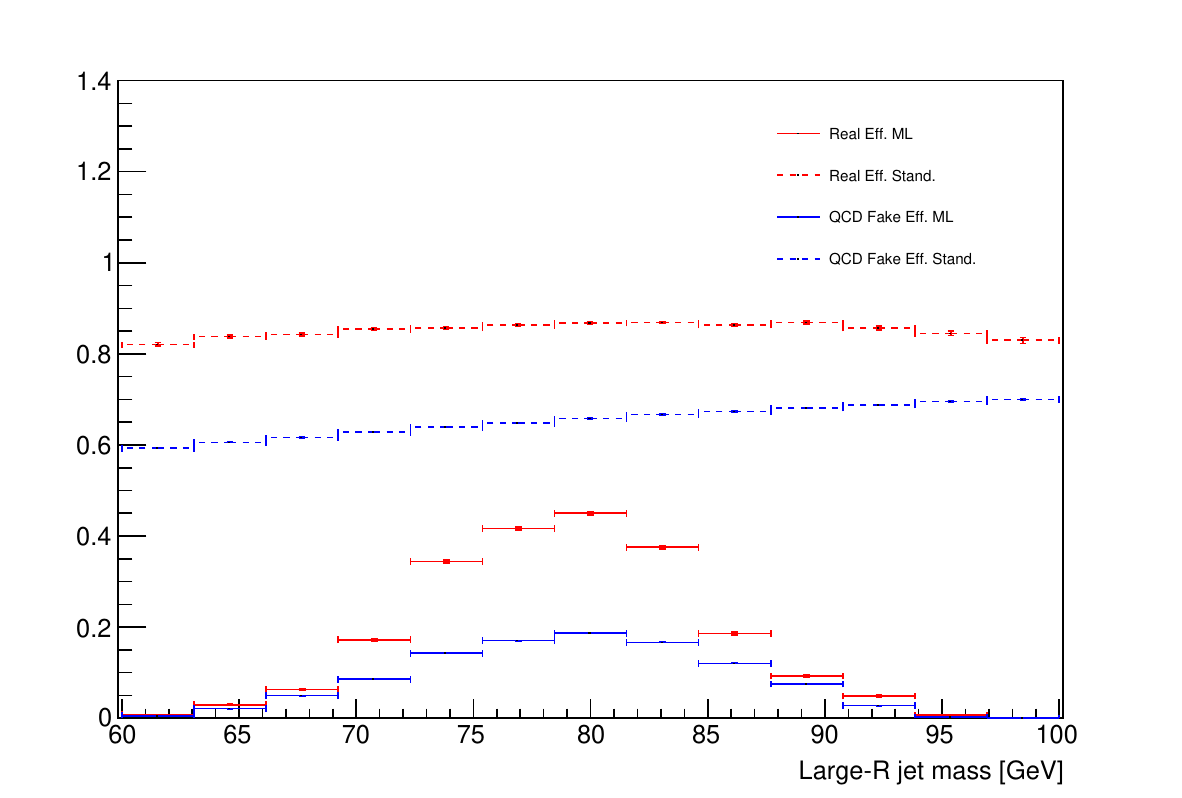}
        \caption{SM $t\bar{t}$.} \label{fig:1b_eff_smtt_w}
      \end{subfigure}%
      \hspace*{\fill}   
      \\
      \begin{subfigure}{0.48\textwidth}
        \includegraphics[width=\linewidth]{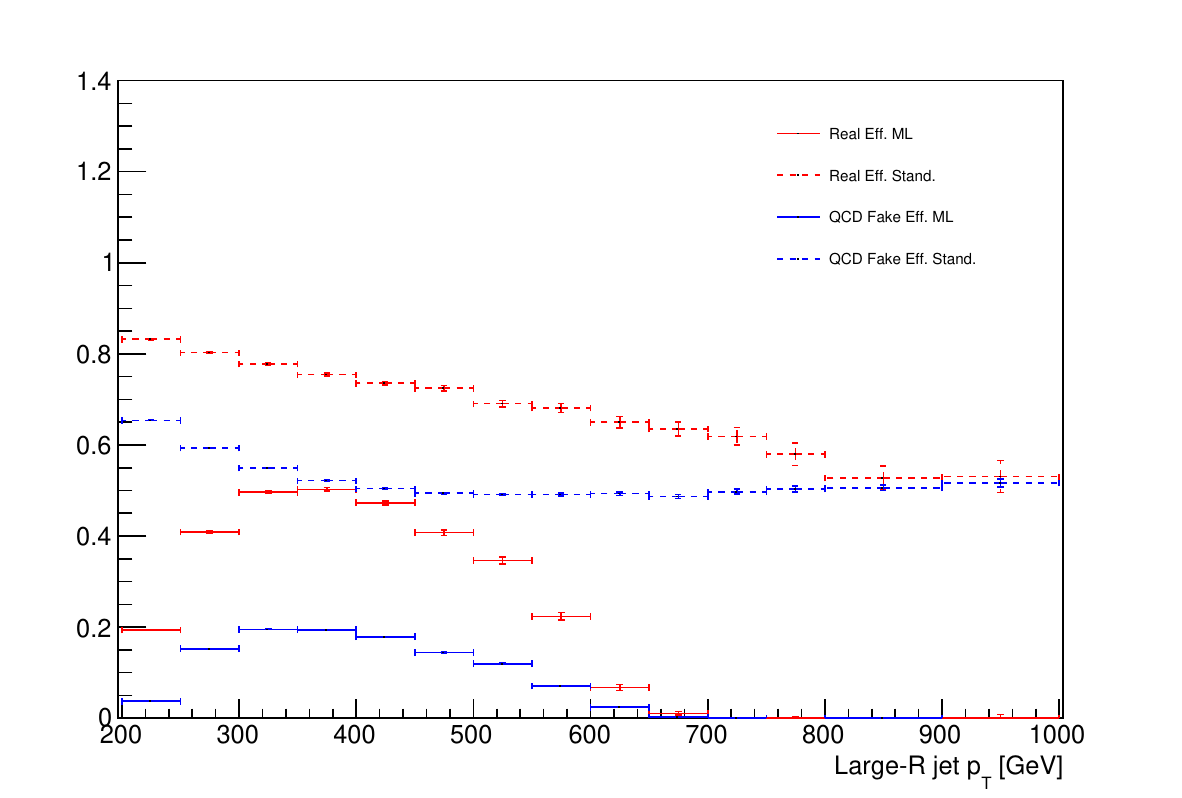}
        \caption{SM $t\bar{t}t\bar{t}$.} \label{fig:1a_eff_smtttt_w}
      \end{subfigure}%
      \hspace*{\fill}   
      \begin{subfigure}{0.48\textwidth}
        \includegraphics[width=\linewidth]{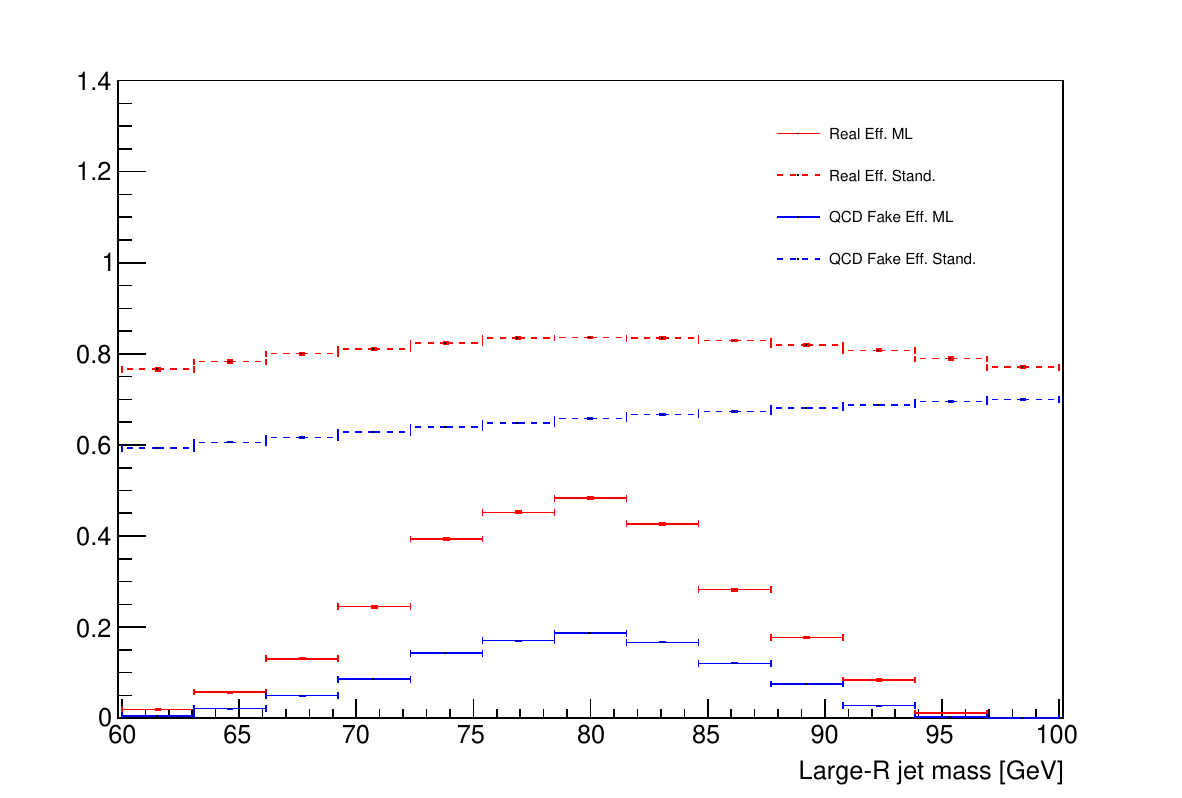}
        \caption{SM $t\bar{t}t\bar{t}$.} \label{fig:1b_eff_smtttt_w}
      \end{subfigure}%
      \hspace*{\fill}   
      \\
      \begin{subfigure}{0.48\textwidth}
        \includegraphics[width=\linewidth]{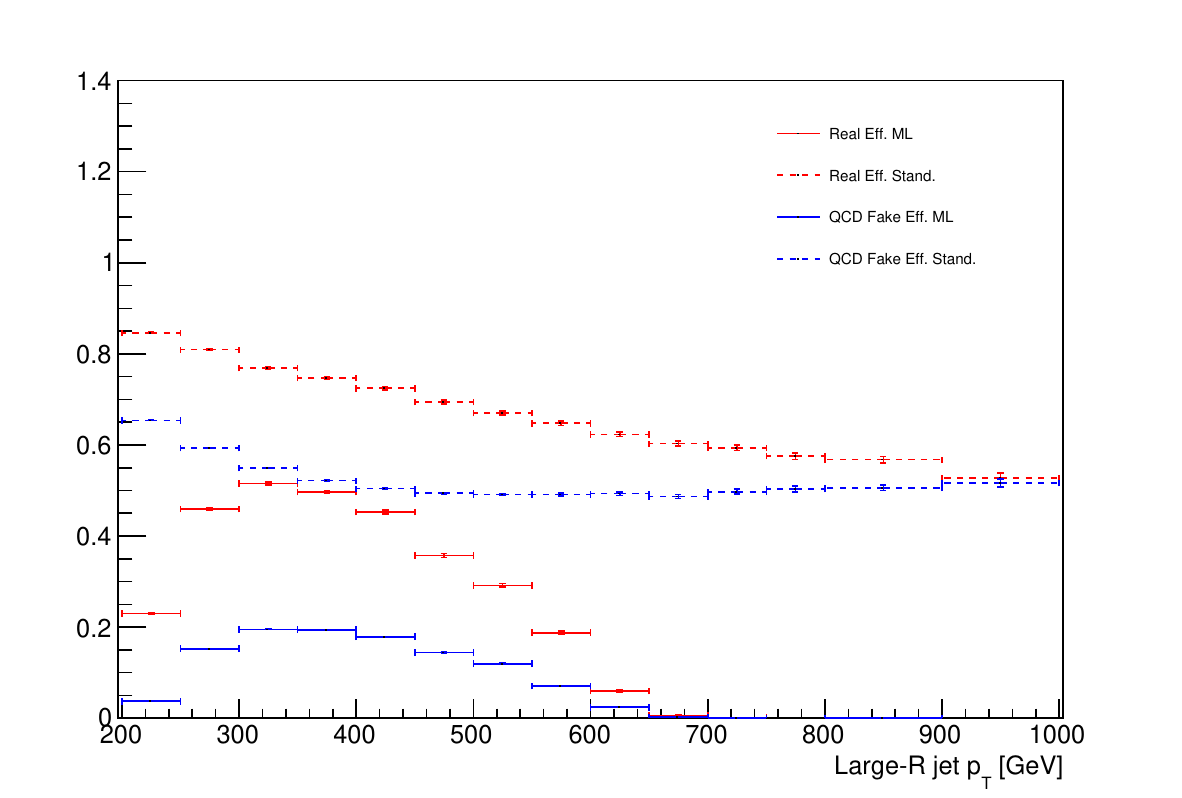}
        \caption{BSM $t\bar{t}y_{0}\rightarrow t\bar{t}t\bar{t}$.} \label{fig:1a_eff_y0tttt_w}
      \end{subfigure}%
      \hspace*{\fill}   
      \begin{subfigure}{0.48\textwidth}
        \includegraphics[width=\linewidth]{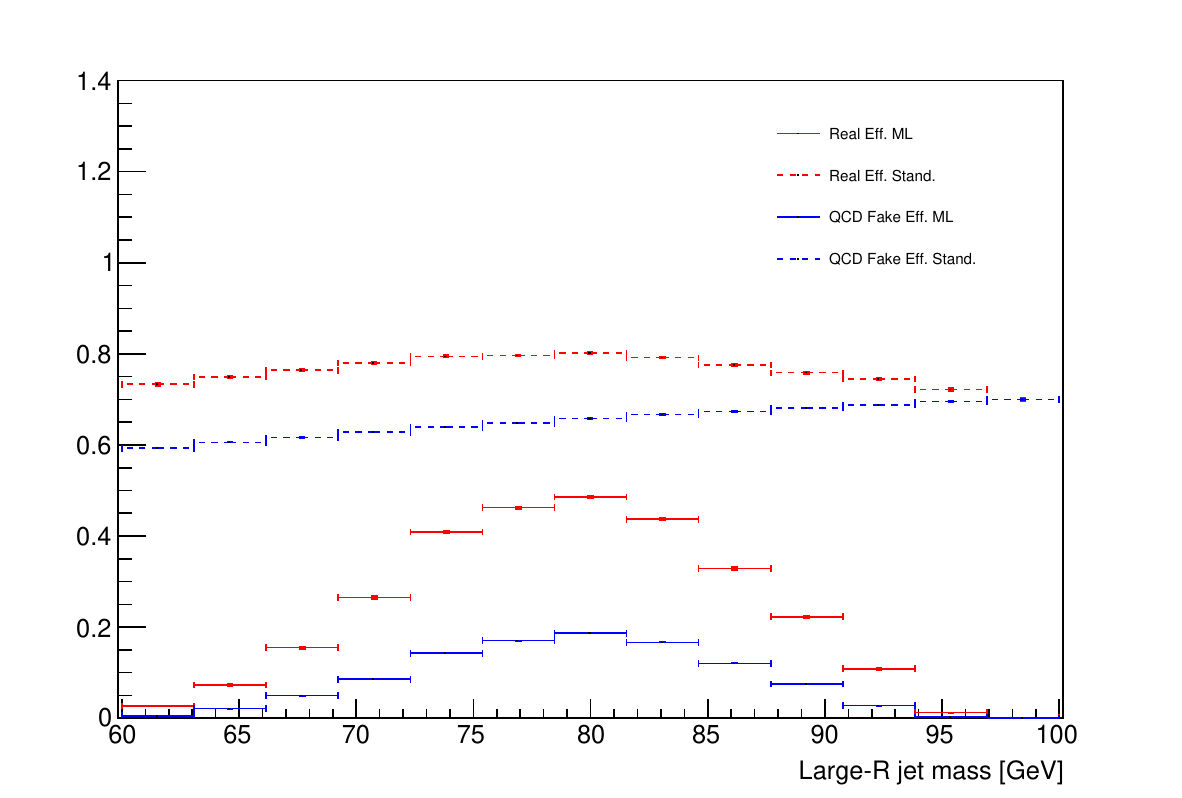}
        \caption{BSM $t\bar{t}y_{0}\rightarrow t\bar{t}t\bar{t}$.} \label{fig:1b_eff_y0tttt_w}
      \end{subfigure}%
      \hspace*{\fill}   

      \caption{$W$-tagging real efficiencies (red) and mistagging rates (blue) using cut-based (dashed lines) and ML-based (solid lines) of 
      (a), (b) SM $t\bar{t}$, (c), (d) SM $t\bar{t}t\bar{t}$, and (e), (f) BSM $t\bar{t}y_{0}\rightarrow t\bar{t}t\bar{t}$ 
      as a function of jet $p_T$ (left) and jet mass (right). 
      The mistagging rates were applied on QCD background samples.}
    \label{fig:eff_smtt_w}
\end{figure}

\begin{figure}[!h]
    \centering
    \begin{subfigure}{0.48\textwidth}
        \includegraphics[width=\linewidth]{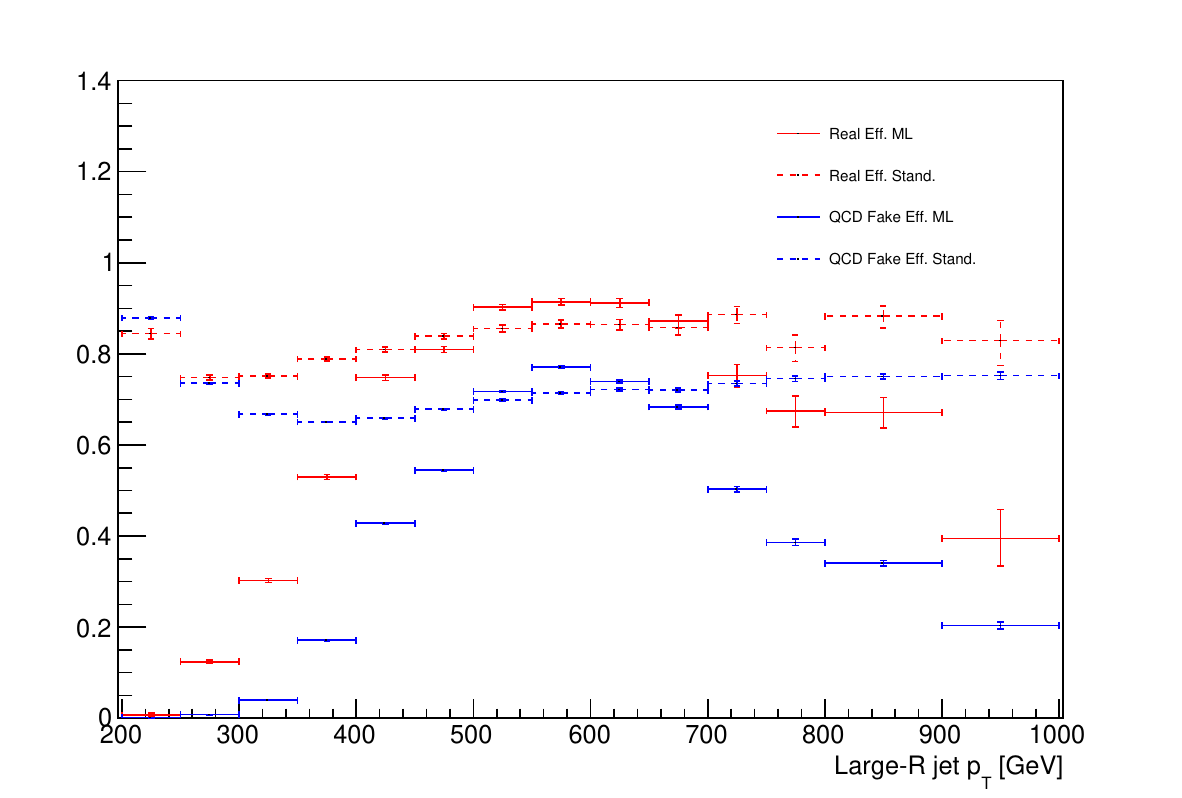}
        \caption{SM $t\bar{t}$.} \label{fig:1a_eff_smtt_top}
      \end{subfigure}%
      \hspace*{\fill}   
      \begin{subfigure}{0.48\textwidth}
        \includegraphics[width=\linewidth]{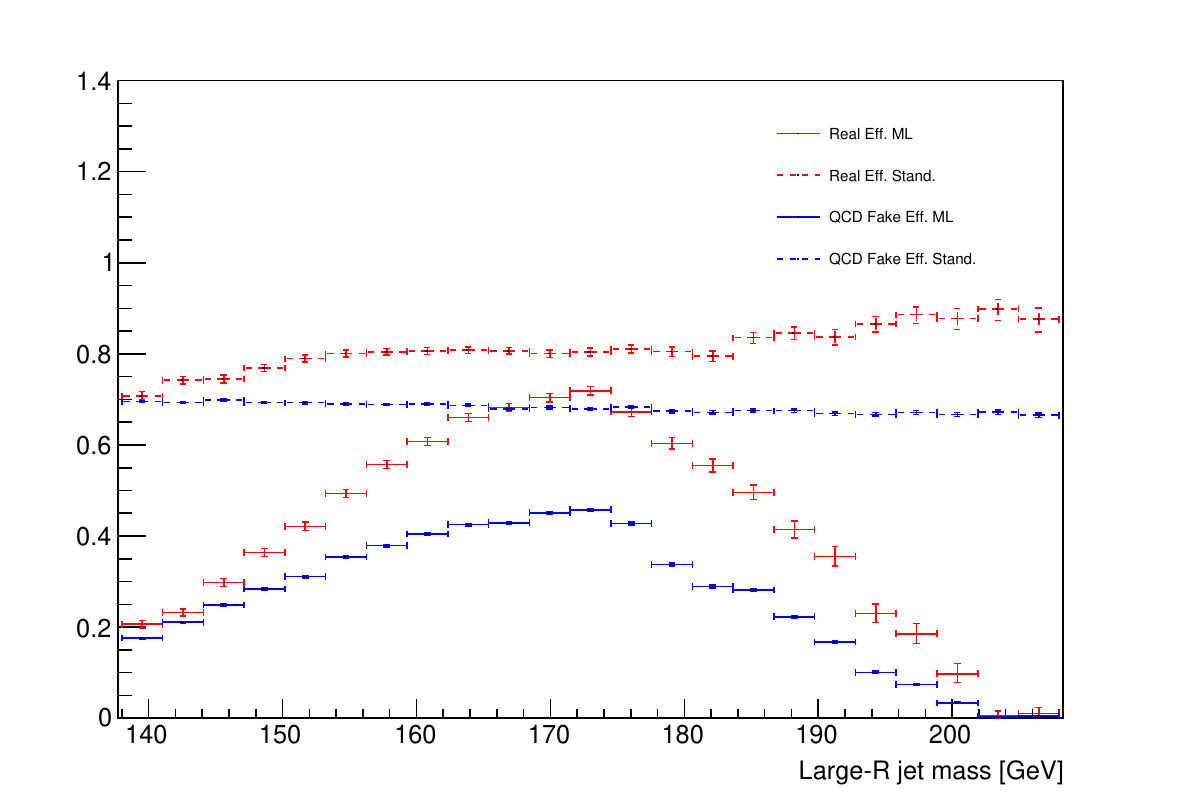}
        \caption{SM $t\bar{t}$.} \label{fig:1b_eff_smtt_top}
      \end{subfigure}%
      \hspace*{\fill}   
      \\
      \begin{subfigure}{0.48\textwidth}
        \includegraphics[width=\linewidth]{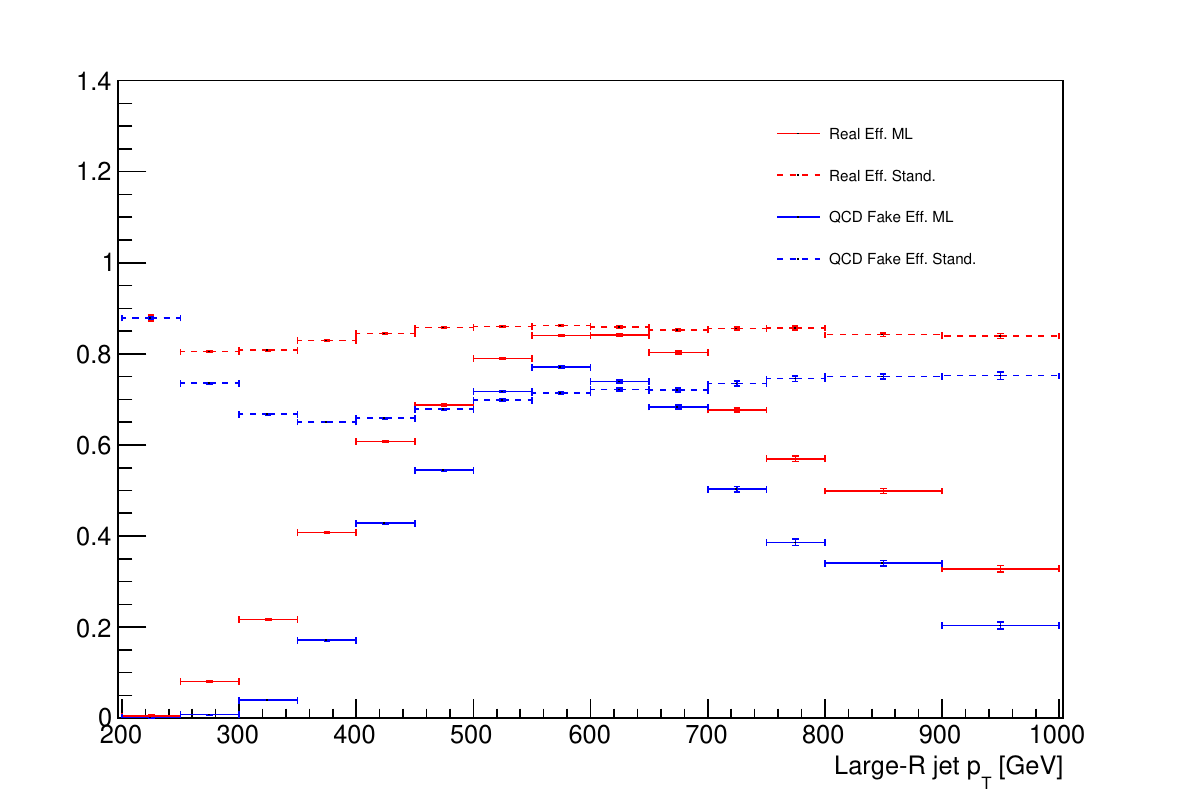}
        \caption{SM $t\bar{t}t\bar{t}$.}  \label{fig:1a_eff_smtttt_top}
      \end{subfigure}%
      \hspace*{\fill}   
      \begin{subfigure}{0.48\textwidth}
        \includegraphics[width=\linewidth]{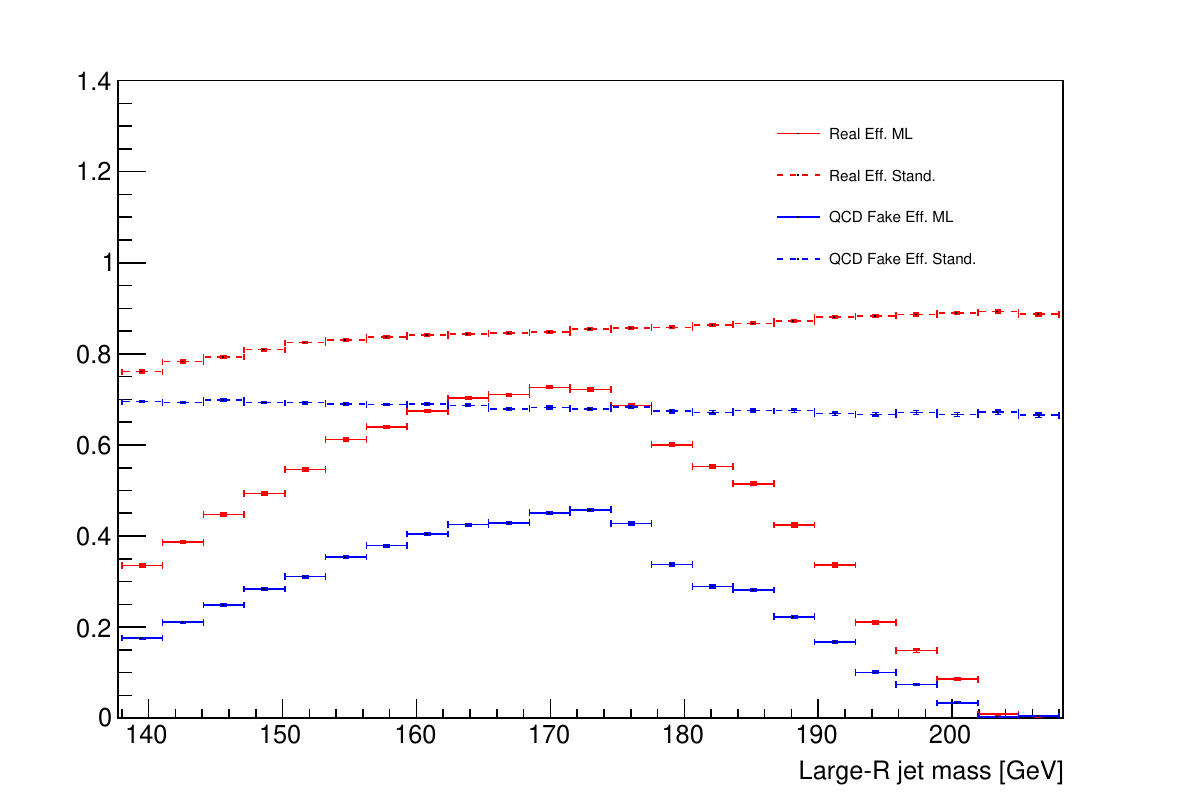}
        \caption{SM $t\bar{t}t\bar{t}$.} \label{fig:1b_eff_smtttt_top}
      \end{subfigure}%
      \hspace*{\fill}   
      \\
      \begin{subfigure}{0.48\textwidth}
        \includegraphics[width=\linewidth]{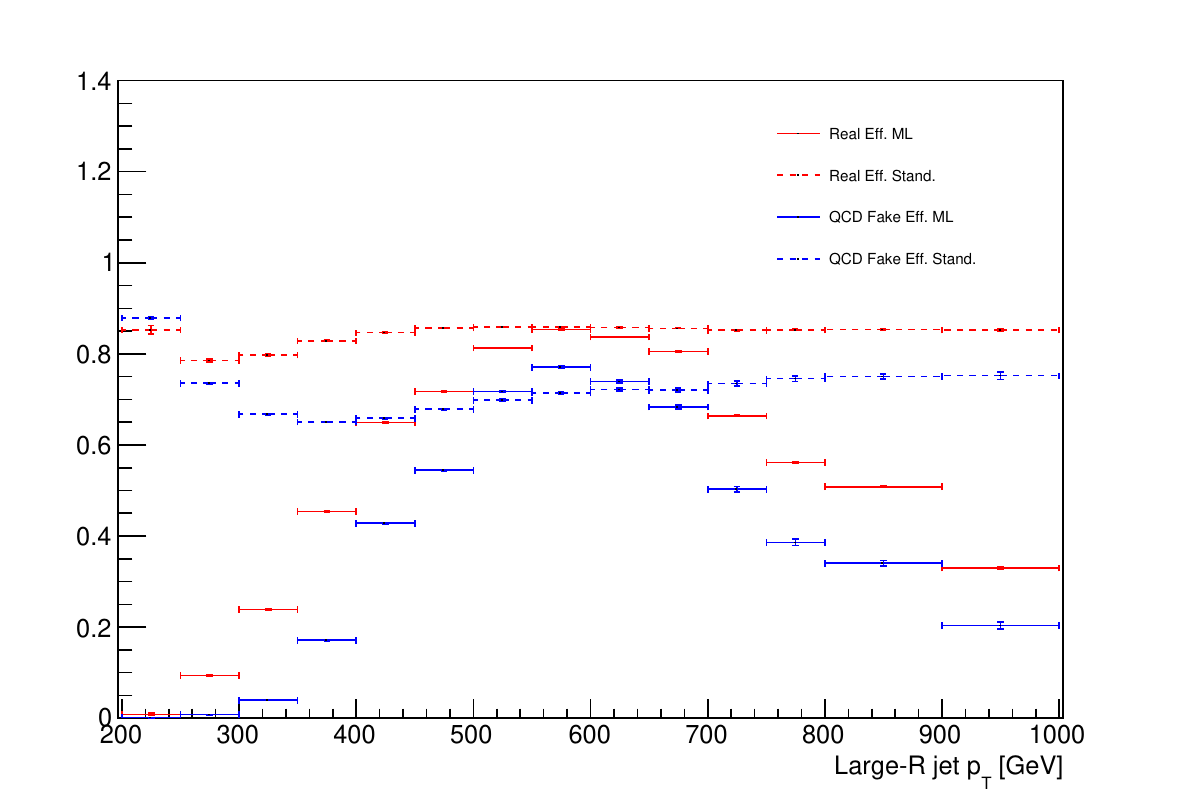}
        \caption{BSM $t\bar{t}y_{0}\rightarrow t\bar{t}t\bar{t}$.} \label{fig:1a_eff_y0tttt_top}
      \end{subfigure}%
      \hspace*{\fill}   
      \begin{subfigure}{0.48\textwidth}
        \includegraphics[width=\linewidth]{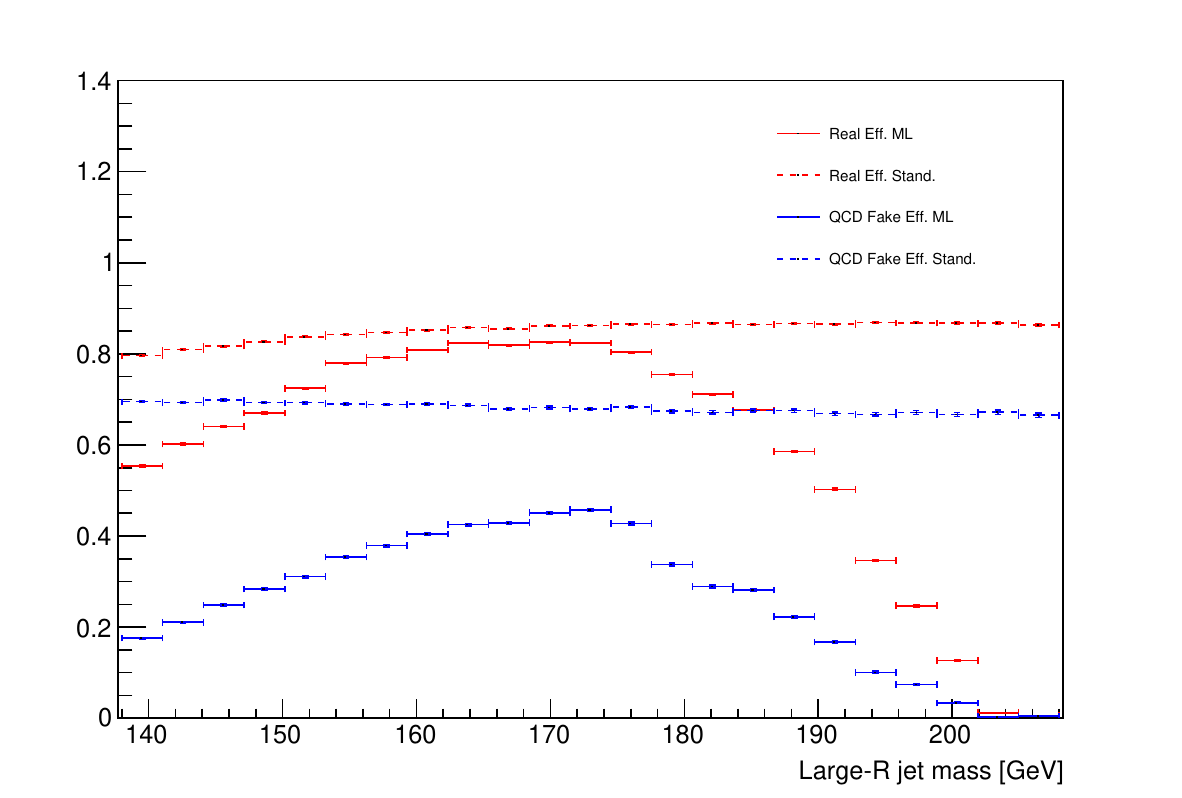}
        \caption{BSM $t\bar{t}y_{0}\rightarrow t\bar{t}t\bar{t}$.} \label{fig:1b_eff_y0tttt_top}
      \end{subfigure}%
      \hspace*{\fill}   

    \caption{Top tagging real efficiencies (red) and mistagging rates (blue) using cut-based (dashed lines) and ML-based (solid lines) of 
    (a), (b) SM $t\bar{t}$, (c), (d) SM $t\bar{t}t\bar{t}$, and (e), (f) BSM $t\bar{t}y_{0}\rightarrow t\bar{t}t\bar{t}$ 
    as a function of jet $p_T$ (left) and jet mass (right). 
    The mistagging rates were applied on QCD background samples.}
    \label{fig:eff_smtt_top}
\end{figure}

The top tagging (Figure~\ref{fig:eff_smtt_top}) and $W$-tagging (Figure~\ref{fig:eff_smtt_w}) efficiencies for cut-based 
(dashed lines) and ML-based (solid lines) are shown for SM $t\bar{t}$ (Figures~\ref{fig:1a_eff_smtt_top},
~\ref{fig:1b_eff_smtt_top},~\ref{fig:1a_eff_smtt_w},~\ref{fig:1b_eff_smtt_w}), 
SM $t\bar{t}t\bar{t}$ (Figures~\ref{fig:1a_eff_smtttt_top},~\ref{fig:1b_eff_smtttt_top},~\ref{fig:1a_eff_smtttt_w},~\ref{fig:1b_eff_smtttt_w}), 
and BSM $t\bar{t}y_{0}\rightarrow t\bar{t}t\bar{t}$ (Figures~\ref{fig:1a_eff_y0tttt_top},~\ref{fig:1b_eff_y0tttt_top},~\ref{fig:1a_eff_y0tttt_w},
~\ref{fig:1b_eff_y0tttt_w}) production.\par
The real efficiencies of cut-based method in both top and $W$-tagging are about 80\%, mostly flat, but also having high 
mistagging rates of about 65-70\%. ML-based method exhibits only slightly lower efficiencies in central mass regions, but the mistagging rates are much suppressed compared to cut-based method, especially in off-peak mass ranges, which helps to make the mass peaks pro pronounced in mass spectra. See 
Figures~\ref{fig:1a_eff_smtt_w},~\ref{fig:1b_eff_y0tttt_w} for $W$-tagging.
and Figures~\ref{fig:1b_eff_y0tttt_top} for BSM model of top tagging. 
For further detail, confusion matricies are shown in the Appendix in~Figure~\ref{fig:mat}.

\clearpage
\subsection{Jet mass spectra}        

\begin{figure}[!h]
  \centering
  \begin{subfigure}{0.9\textwidth}
      \includegraphics[width=\linewidth]{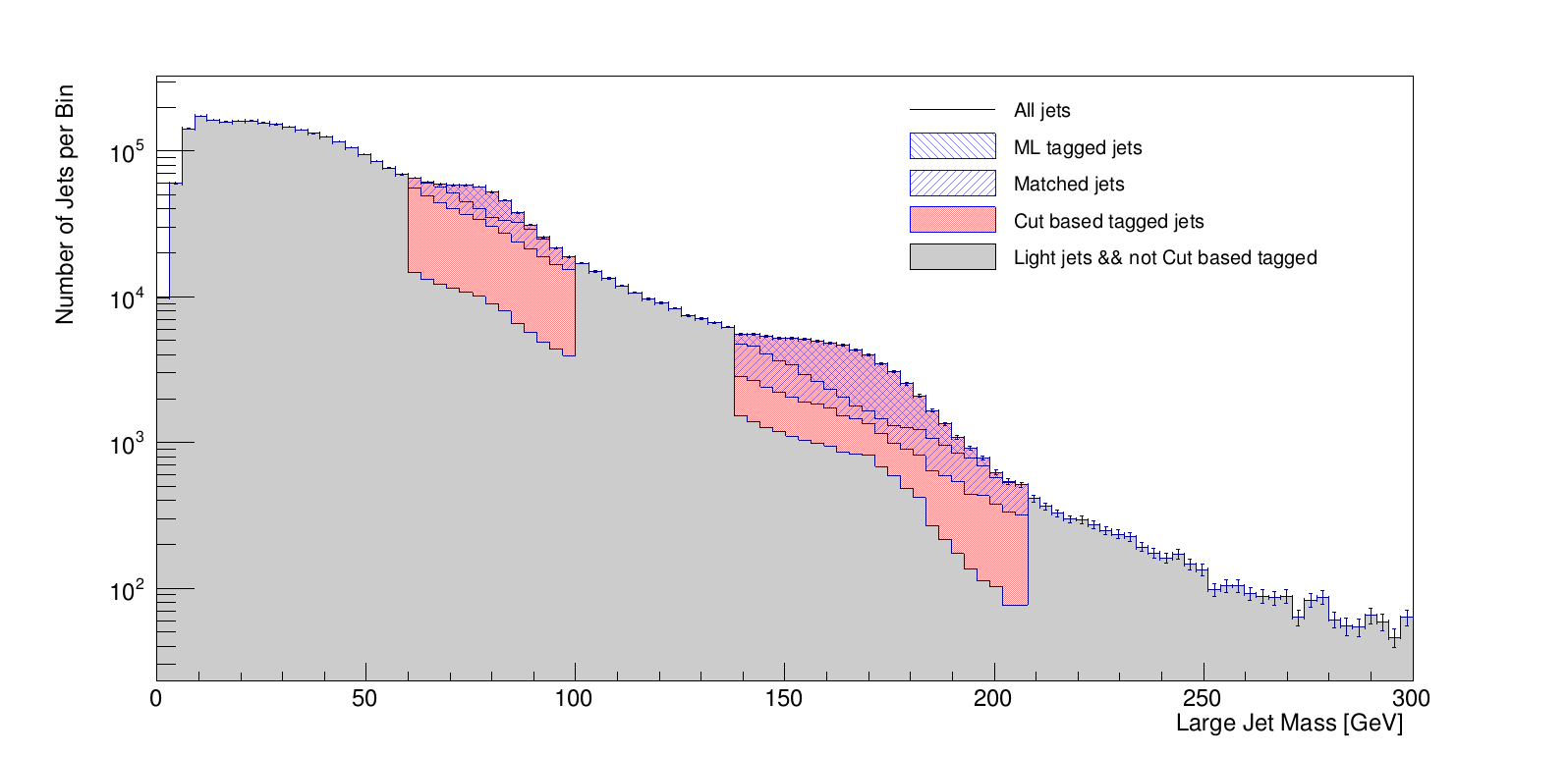}
      \caption{SM $t\bar{t}$.} \label{fig:1a_peak}
    \end{subfigure}%
    \\ 
    \begin{subfigure}{0.9\textwidth}
      \includegraphics[width=\linewidth]{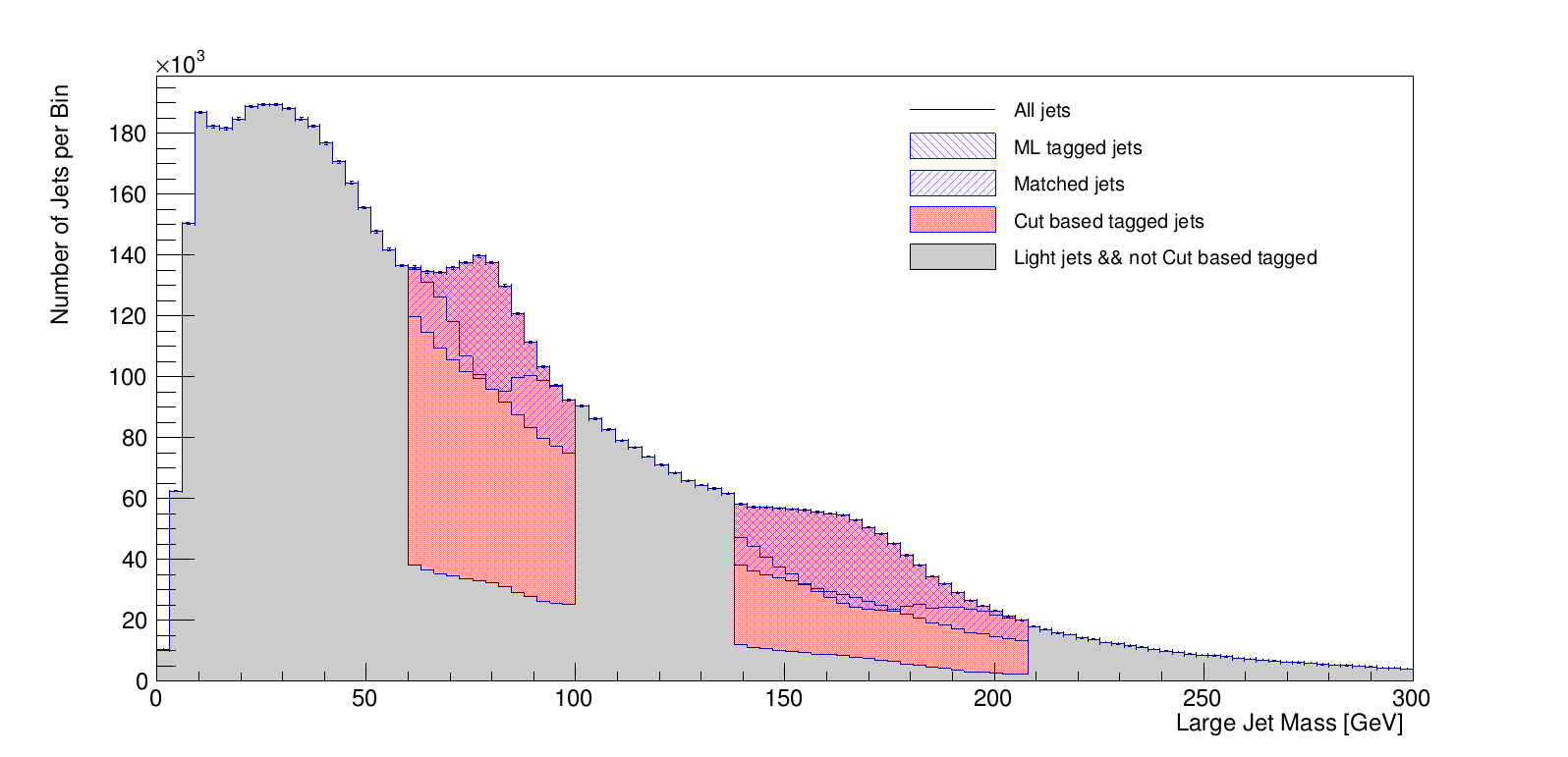}
      \caption{SM $t\bar{t}t\bar{t}$.} \label{fig:1b_peak}
    \end{subfigure}%
    \\ 
    \begin{subfigure}{0.9\textwidth}
      \includegraphics[width=\linewidth]{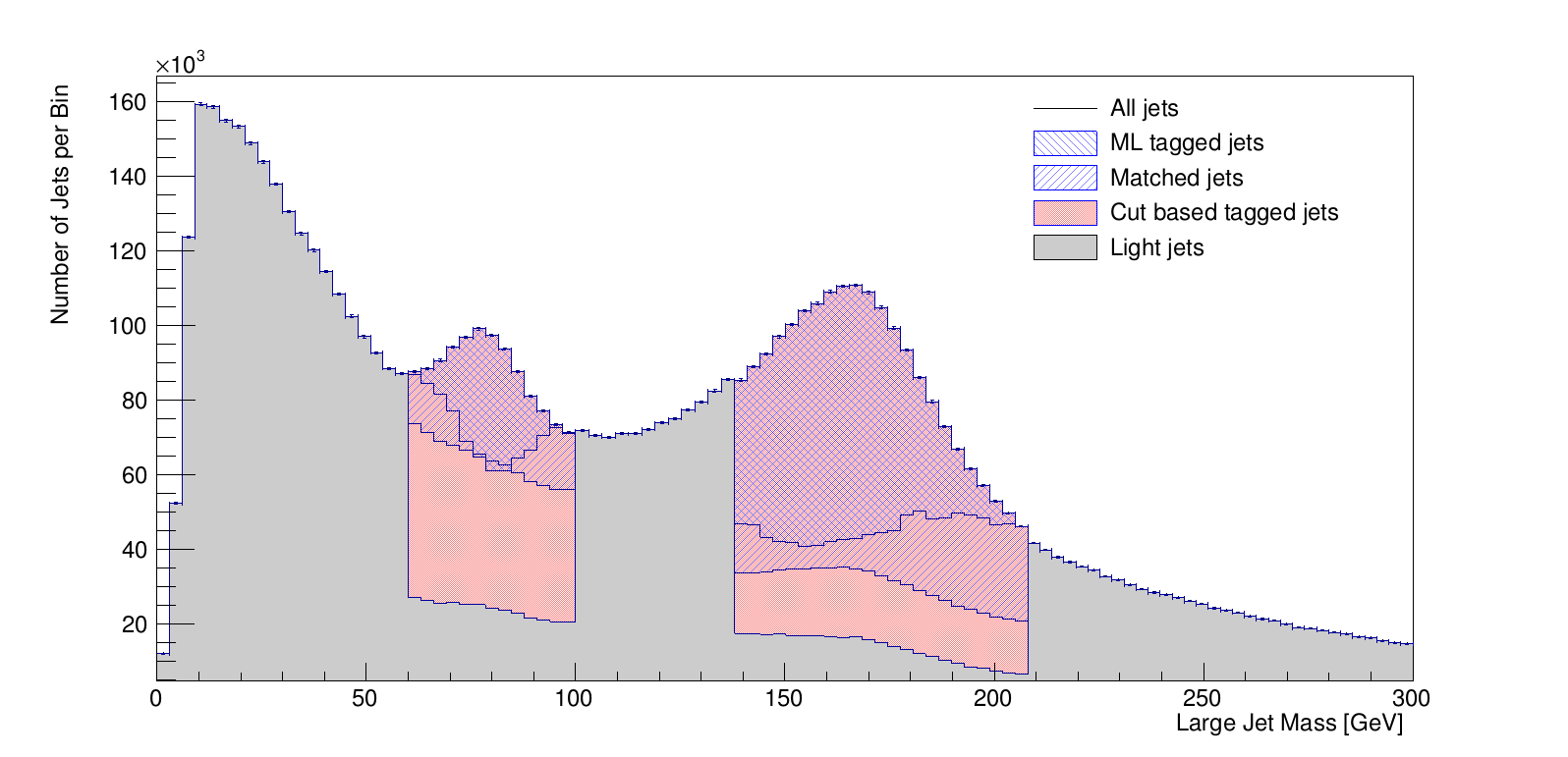}
      \caption{BSM $t\bar{t}y_{0}\rightarrow t\bar{t}t\bar{t}$.} \label{fig:1c_peak}
    \end{subfigure}
  
    \caption{The jet mass spectra of (a) SM $t\bar{t}$, (b) SM $t\bar{t}t\bar{t}$, and (c) BSM $t\bar{t}y_{0}\rightarrow t\bar{t}t\bar{t}$.
    The pink area stands for cut-based, hatched area for ML-based, and semi-hatched area for the "true" tagging.}
  \label{fig:peaks}
\end{figure}

The spectra of the large-$R$ jet mass could help understand whether the "true" jet label based on jet angular matching to top and $W$ generated particles is performs as expected.
Figure~\ref{fig:peaks} presents large jet mass spectra of SM $t\bar{t}$ (Figure~\ref{fig:1a_peak}), SM $t\bar{t}t\bar{t}$ 
(Figure~\ref{fig:1b_peak}), and BSM $t\bar{t}y_{0}\rightarrow t\bar{t}t\bar{t}$ (Figure~\ref{fig:1c_peak}) 
with areas highlighted for cut-based (pink area), ML-based (hatched area) 
tagging, also mostly in between (semi-hatched area) stands for those matched to top or $W$ (defining the "true" labels).\par
While cut-based method tags a large portion light jets as top or $W$ jets, the ML-based method tags top and $W$ jets 
closer the the "true" labeled jets, especially around the means of the top and $W$ mass peaks.
However, the performance of "true" jets definition on the samples with a large number of top and $W$ particles 
and additional number of jets is not ideal since a non-negligible part of light jets still passes the jet matching algorithm.

\clearpage
\subsection{Spectrum of invariant mass of two jets}   

\begin{figure}[!h]
  \centering
  \begin{subfigure}{0.95\textwidth}
      \includegraphics[width=\linewidth]{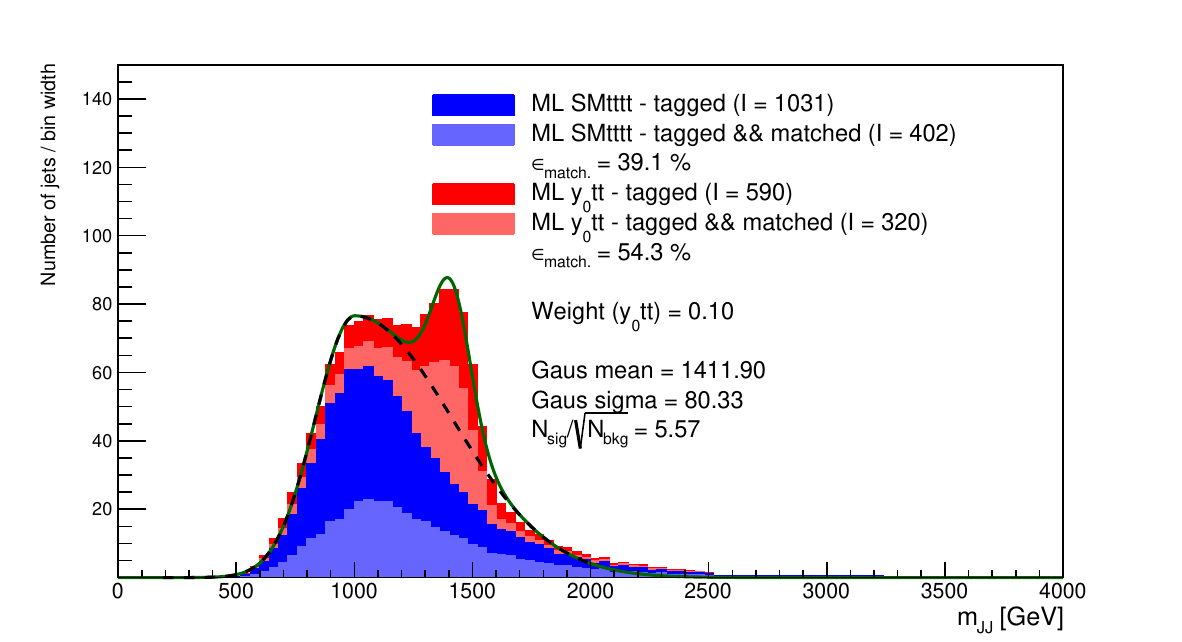}
      \caption{ML-based method} \label{fig:1a_peak2}
    \end{subfigure}%
    \\ 
    \begin{subfigure}{0.95\textwidth}
      \includegraphics[width=\linewidth]{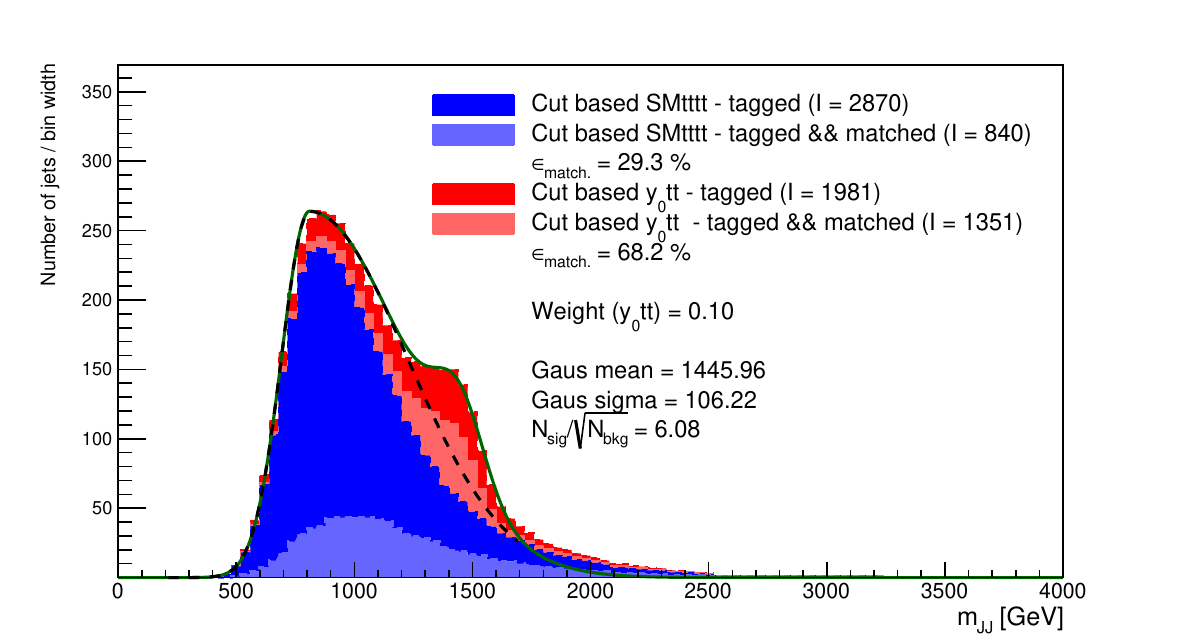}
      \caption{Cut-based method.} \label{fig:1b_peak2}
    \end{subfigure}%
  \caption{Invariant mass of two $t$tagged jets (all possible combinations) for the process of SM $t\bar{t}t\bar{t}$ (blue area) 
  representing background process with the stacked signal process $t\bar{t}y_{0}\rightarrow t\bar{t}t\bar{t}$ (red area) scaled 
  to its 10\%. The light red and blue areas show tagged and matched jets to highlight the tagging efficiencies. The background 
  fit is given by black line using Bifurcated Gaussian and green line is the Gaussian signal fit.}
  \label{fig:peaks2}
\end{figure}

This section describes a performance of the developed tagging algorithms on simulations involving a BSM signal. 
Figure~\ref{fig:peaks2} represents stacked histograms of the dijet 
invariant mass where both jets were tagged as $t$-jets, with all possible jet combinations used, assuming a SM $t\bar{t}t\bar{t}$ as a background process (blue area) and an additiona BSM signal process $t\bar{t}y_{0}\rightarrow t\bar{t}t\bar{t}$ (red area) scaled by an arbitrary factor of $0.1$. 

The blue and red colors are divided into lighter and darker to show the tagging efficiencies. The ML-based method performance is shown in Figure~\ref{fig:1a_peak2}, while cut-based method in Figure~\ref{fig:1b_peak2}.\par

We perform an exercise of finding a signal peak over a falling background by performing a background fit using a Bifurcated Gaussian function and an additional Gaussian function for the the signal peak modelling. The signal significance calculated based on the fitted areas turns out to be slightly higher for cut-based method ($\mathrm{N}_{\mathrm{sig}}/\sqrt{\mathrm{N}_{\mathrm{bkg}}} \doteq 6.1$) compare to ML-based method ($\mathrm{N}_{\mathrm{sig}}/\sqrt{\mathrm{N}_{\mathrm{bkg}}} \doteq 5.6$).
On the other hand the signal peak mass resolution (standard deviation of signal Gaussian fit) is smaller in case of the ML-based method, $\sigma \doteq 80$~GeV compare the the cut-based method, $\sigma \doteq 106$~GeV.

\section{Conclusions}

This study demonstrates the power of machine learning (ML) techniques, particularly Gradient Boosting Classifiers (GBC) and Multi-Layer Perceptrons (MLP), in tagging hadronic jets originating from top quarks and $W$ bosons, copmared to classical cut-based techniques using the same input variables.
By leveraging classical subjettiness variables within a parameterized detector simulation framework, the presented ML-based approach provides a significant improvement in mistagging rates compared to traditional cut-based methods, especially in the context of complex hadronic environments such as the four-top quark final state.

The lower mistagging rates achieved by the ML models are particularly promising for reducing multijet backgrounds in current or future high-energy physics experiments, which is crucial for identifying rare signals such as those from Beyond Standard Model (BSM) processes. The presented simple ML approach does come with a trade-off in slightly lower real tagging efficiencies, which however is not the case of more developped techniques already used in HEP experients. However, one of our goals was to compare the ML and cut-based approaches using the same inputs.

When comparing the ML-based and cut-based methods, a key metric is the significance of signal detection. 
In this study, the cut-based method yielded a slightly higher significance compared to the ML-based method. This difference suggests that while the presented ML-based method excels in reducing false positives, the cut-based method might still be more effective in scenarios where maximizing the raw signal strength is critical, but applicable mostly in regions of large signal-to-background ration which is not often the case.

But clearly, the observed signal mass peak resolution of a di-top resonance was notably smaller for the ML-based method compared to the cut-based method, indicating that the ML-based method provides a tighter and more accurate representation of the signal 
which is crucial for precise mass measurements or for distinguishing closely spaced signals or in areas where a signal peak is close to a kinematic peak.




\section{Acknowledgments}
Authors would like to thank the Czech Science Foundation projects GAČR 23-07110S for the support of this work.


\bibliography{main}{}
\bibliographystyle{unsrt}

\section{Appendix}
\subsection{Performance of ML-based algorithm}\label{app:perf}
In this section detailed view on performace of ML-based algorithms is given in the following tables. 

\begin{itemize}
  \item[]
  \begin{center}
  \begin{tabular}{|c|c|c|c|}
  \hline
  {\bf Measures} & {\bf Training data set} & {\bf First testing data set} & {\bf Second testing data set}\\
  \hline
  {\bf Accuracy} & 68.2\% & 68.1\% & 62.3\% \\
  {\bf Precision} & 58.1\% & 57.9\% & 57.5\% \\
  {\bf Recall} & 35.1\% & 35.1\% & 42.8\%\\
  {\bf FPR} & 13.8\% & 13.9\% & 23.4\% \\
  \hline
  \end{tabular}
  \captionof{table}{Performance metrics of GBC model for $W$ tagging}
  \end{center}
  
  \item[]
  \begin{center}
  \begin{tabular}{|c|c|c|c|}
  \hline
  {\bf Measures} & {\bf Training data set} & {\bf First testing data set} & {\bf Second testing data set}\\
  \hline
  {\bf Accuracy} & 68\% & 67.9\% & 62.2\% \\
  {\bf Precision} & 57.6\% & 57.3\% & 57.4\% \\
  {\bf Recall} & 35.2\% & 35.2\% & 42.3\% \\
  {\bf FPR} & 14.2\% & 14.3\% & 23.1\% \\
  \hline
  \end{tabular}
  \captionof{table}{Performance metrics of MLP model for $W$ tagging}
  \end{center}
\end{itemize}
  
\begin{itemize}
  \item[]
  \begin{center}
  \begin{tabular}{|c|c|c|c|}
  \hline
  {\bf Measures} & {\bf Training data set} & {\bf First testing data set} & {\bf Second testing data set}\\
  \hline
  {\bf Accuracy} & 66.8\% & 66.2\% & 62.2\% \\
  {\bf Precision} & 84.8\% & 84.3\% & 72.8\% \\
  {\bf Recall} & 69.7\% & 69.4\% & 61.8\%\\
  {\bf FPR} & 43.4\% & 44.7\% & 37.2\%\\
  \hline
  \end{tabular}
  \captionof{table}{Performance metrics of GBC model with random undersampling for $t$-tagging}
  \end{center}
  
  \item[]
  \begin{center}
  \begin{tabular}{|c|c|c|c|}
  \hline
  {\bf Measures} & {\bf Training data set} & {\bf First testing data set} & {\bf Second testing data set}\\
  \hline
  {\bf Accuracy} & 65.4\% & 65.3\% & 61.9\% \\
  {\bf Precision} & 84.7\% & 84.6\% & 73.1\% \\
  {\bf Recall} & 67.6\% & 67.7\% & 60.5\%\\
  {\bf FPR} & 42.4\% & 42.9\% & 36.0\% \\
  \hline
  \end{tabular}
  \captionof{table}{Performance metrics of MLP model with random undersampling for $t$-tagging}
  \end{center}
  
  \item[]
  \begin{center}
  \begin{tabular}{|c|c|c|c|}
  \hline
  {\bf Measures} & {\bf Training data set} & {\bf First testing data set} & {\bf Second testing data set}\\
  \hline
  {\bf Accuracy} & 58.6\% & 58.4\% & 59.2\%\\
  {\bf Precision} & 84.6\% & 84.4\% & 72\%\\
  {\bf Recall} & 57\% & 57\% & 55.5\%\\
  {\bf FPR} & 35.9\% & 36.5\% & 34.8\% \\
  \hline
  \end{tabular}
  \captionof{table}{Performance metrics of GBC model with cluster centroids for $t$-tagging}
  \end{center}
  
  \item[]
  \begin{center}
  \begin{tabular}{|c|c|c|c|}
  \hline
  {\bf Measures} & {\bf Training data set} & {\bf First testing data set} & {\bf Second testing data set}\\
  \hline
  {\bf Accuracy} & 59.8\% & 59.9\% & 59.9\%\\
  {\bf Precision} & 84.4\% & 84.5\% & 72.1\% \\
  {\bf Recall} & 59.1\% & 59.2\% & 57.2\% \\
  {\bf FPR} & 37.9\% & 37.6\% & 35.7\% \\
  \hline
  \end{tabular}
  \captionof{table}{Performance metrics of MLP model with cluster centroids for $t$-tagging}
  \end{center}
  
  \item[]
  \begin{center}
  \begin{tabular}{|c|c|c|c|}
  \hline
  {\bf Measures} & {\bf Training data set} & {\bf First testing data set} & {\bf Second testing data set}\\
  \hline
  {\bf Accuracy} & 51\% & 50.8\% & 51.1\%\\
  {\bf Precision} & 81.8\% & 81.6\% & 68.4\% \\
  {\bf Recall} & 47.3\% & 47.2\% & 38.8\%\\
  {\bf FPR} & 36.5\% & 36.9\% & 29\%\\
  \hline
  \end{tabular}
  \captionof{table}{Performance metrics of GBC model with near miss for $t$-tagging}
  \end{center}
  
  \item[]
  \begin{center}
  \begin{tabular}{|c|c|c|c|}
  \hline
  {\bf Measures} & {\bf Training data set} & {\bf First testing data set} & {\bf Second testing data set}\\
  \hline
  {\bf Accuracy} & 50.1\% & 50.3\% & 50.9\% \\
  {\bf Precision} & 81.3\% & 81.4\% & 68.1\% \\
  {\bf Recall} & 46.4\% & 46.6\% & 38.5\% \\
  {\bf FPR} & 36.9\% & 36.8\% & 29\% \\
  \hline
  \end{tabular}
  \captionof{table}{Performance metrics of MLP model with near miss for $t$-tagging}
  \end{center}
  
  \item[]
  \begin{center}
  \begin{tabular}{|c|c|c|c|}
  \hline
  {\bf Measures} & {\bf Training data set} & {\bf First testing data set} & {\bf Second testing data set}\\
  \hline
  {\bf Accuracy} & 66.1\% & 55.8\% & 56.8\%\\
  {\bf Precision} & 100\% & 86\% & 74.9\% \\
  {\bf Recall} & 56.3\% & 51.7\% & 45\%\\
  {\bf FPR} & 0\% & 30\% & 24.3\% \\
  \hline
  \end{tabular}
  \captionof{table}{Performance metrics of GBC model with repeated edited nearest neighbors for $t$-tagging}
  \end{center}
  
  \item[]
  \begin{center}
  \begin{tabular}{|c|c|c|c|}
  \hline
  {\bf Measures} & {\bf Training data set} & {\bf First testing data set} & {\bf Second testing data set}\\
  \hline
  {\bf Accuracy} & 59\% & 58\% & 57.5\% \\
  {\bf Precision} & 86.4\% & 85.5\% & 74.5\% \\
  {\bf Recall} & 55.9\% & 55.3\% & 47.3\% \\
  {\bf FPR} & 30.4\% & 32.5\% & 26.1\% \\
  \hline
  \end{tabular}
  \captionof{table}{Performance metrics of MLP model with repeated edited nearest neighbors for $t$-tagging}
  \end{center}
\end{itemize}

\subsection{Confusion matricies}
\begin{figure}[!h]
  \centering
  \begin{subfigure}{0.31\textwidth}
      \includegraphics[width=\linewidth]{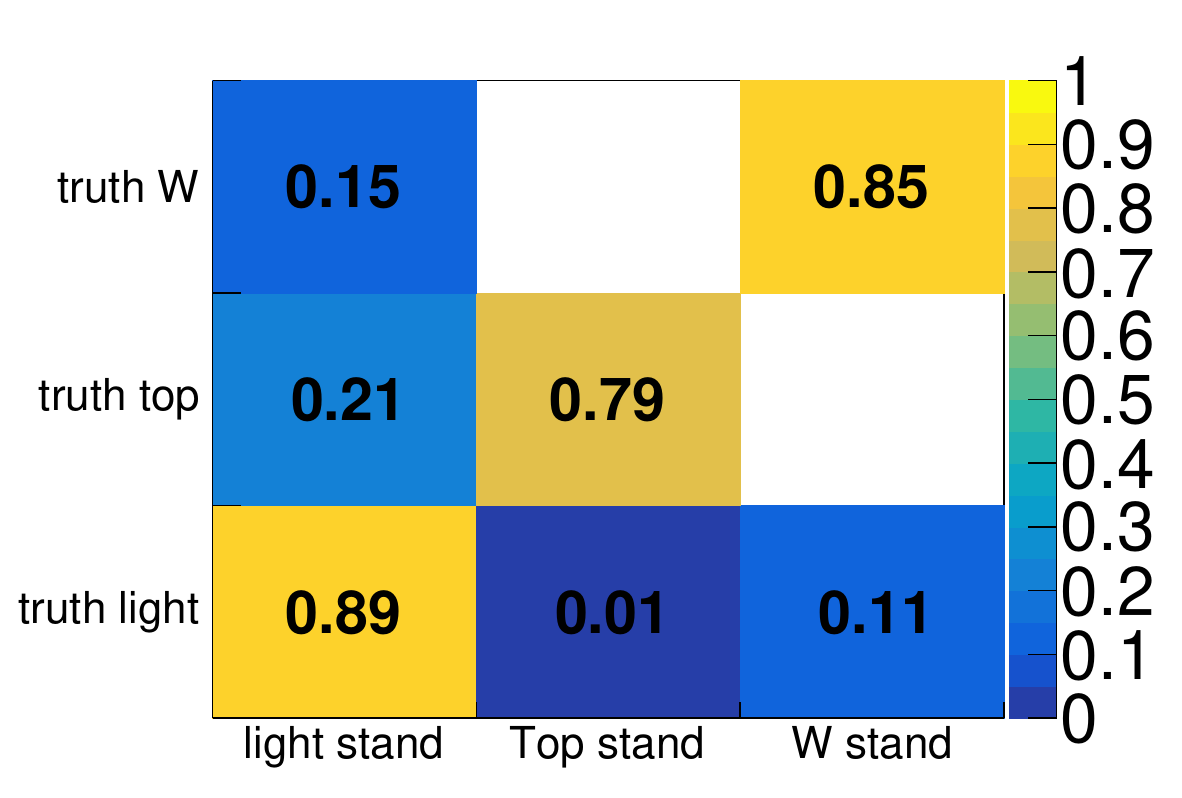}
      \caption{SM $t\bar{t}$.} \label{fig:1a_mat_smtt}
    \end{subfigure}%
    \hspace*{\fill}   
    \begin{subfigure}{0.31\textwidth}
      \includegraphics[width=\linewidth]{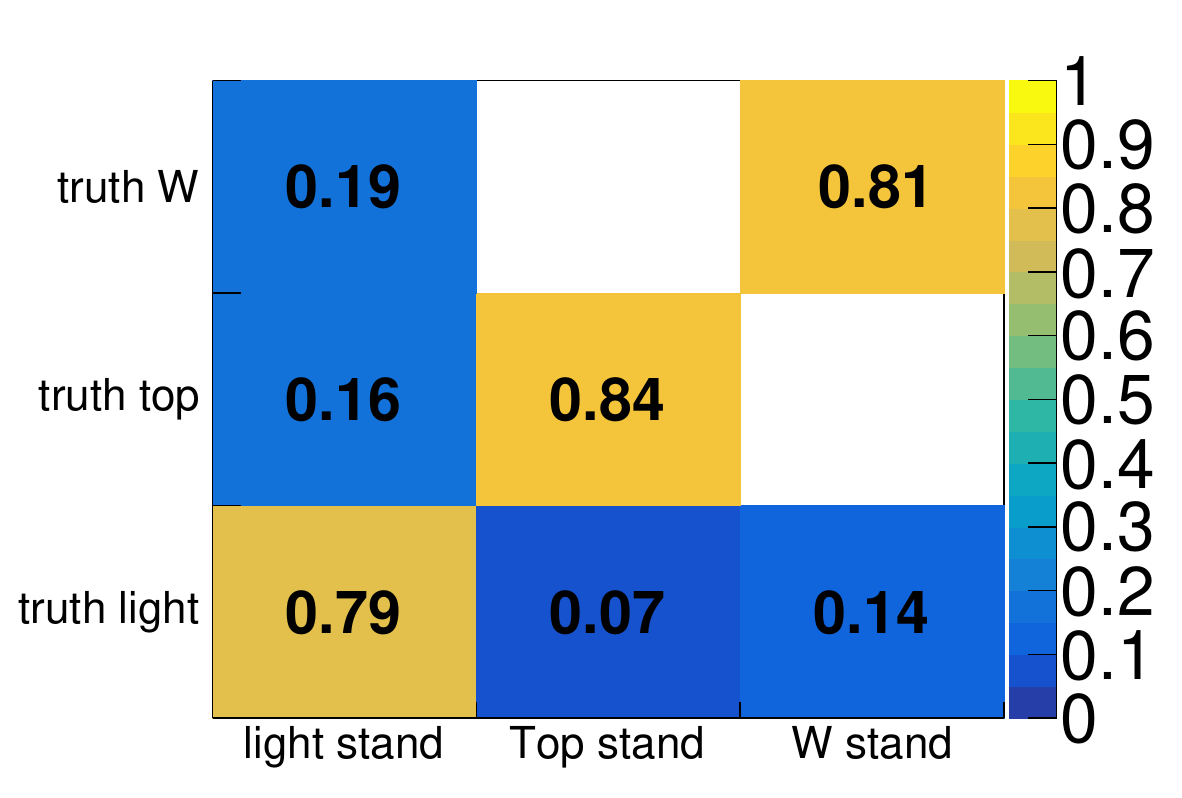}
      \caption{SM $t\bar{t}t\bar{t}$.} \label{fig:1b_mat_smtttt}
    \end{subfigure}%
    \hspace*{\fill}   
    \begin{subfigure}{0.31\textwidth}
      \includegraphics[width=\linewidth]{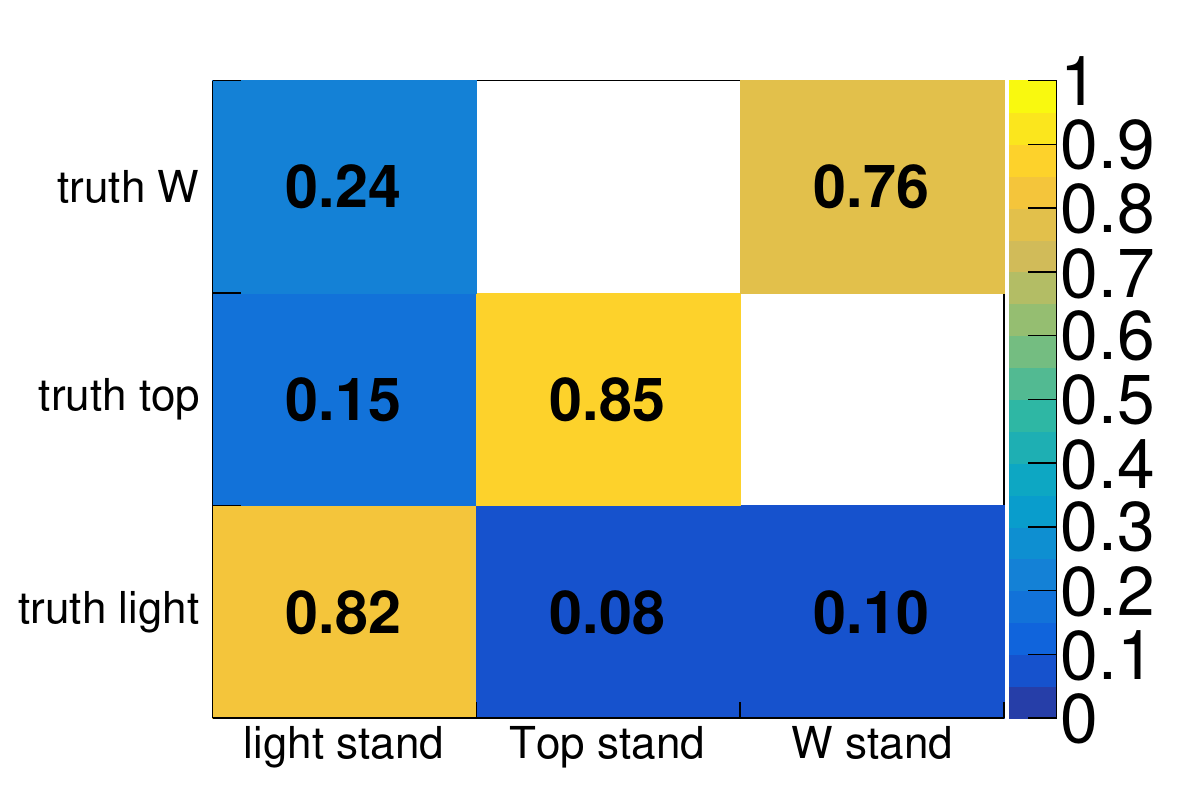}
      \caption{$t\bar{t}y_{0}\rightarrow t\bar{t}t\bar{t}$.} \label{fig:1c_mat_y0}
    \end{subfigure}
    \\
    \begin{subfigure}{0.31\textwidth}
      \includegraphics[width=\linewidth]{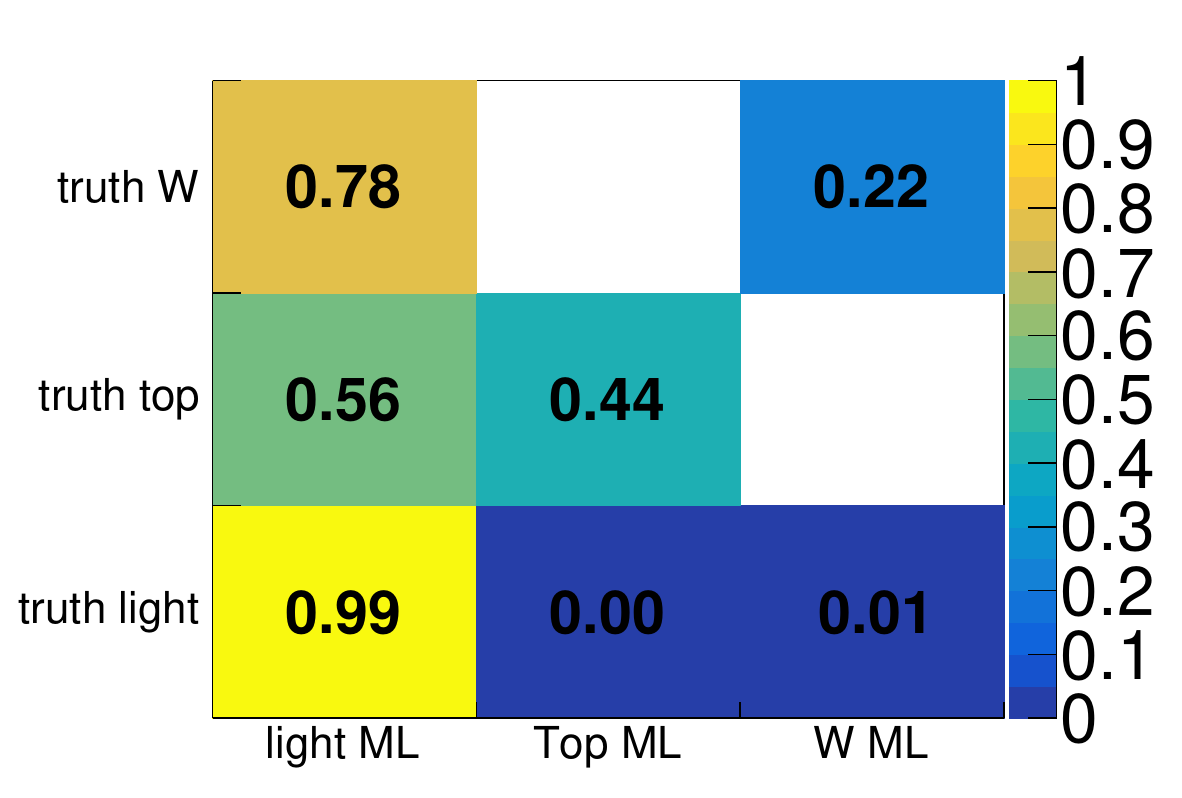}
      \caption{SM $t\bar{t}$.} \label{fig:2a_mat_smtt}
    \end{subfigure}%
    \hspace*{\fill}   
    \begin{subfigure}{0.31\textwidth}
      \includegraphics[width=\linewidth]{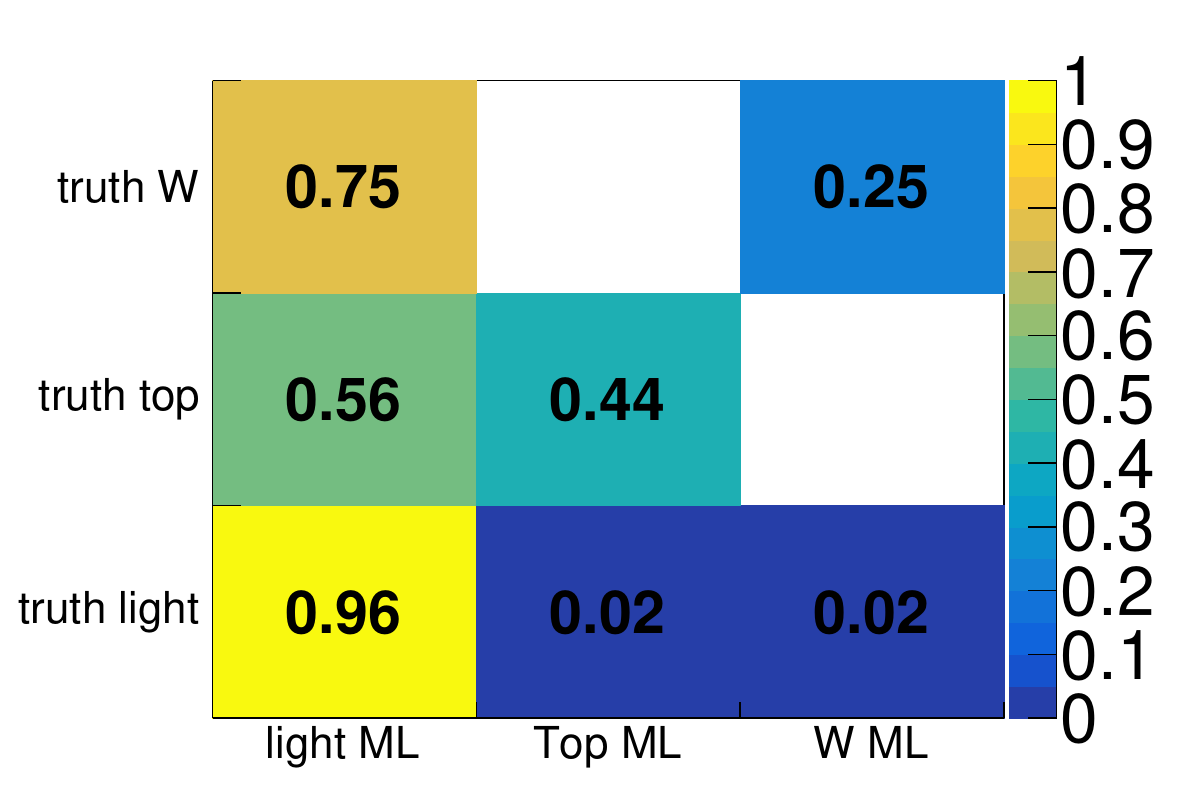}
      \caption{SM $t\bar{t}t\bar{t}$.} \label{fig:2b_mat_smtt}
    \end{subfigure}%
    \hspace*{\fill}   
    \begin{subfigure}{0.31\textwidth}
      \includegraphics[width=\linewidth]{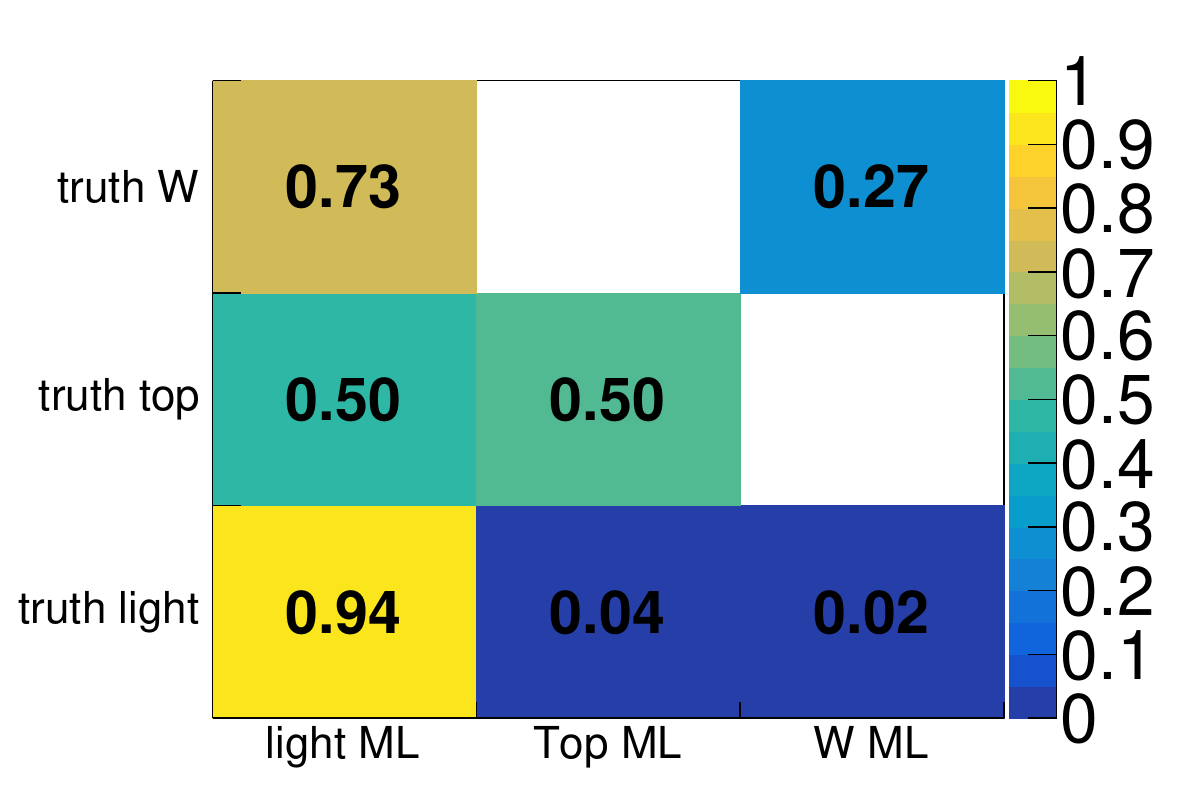}
      \caption{$t\bar{t}y_{0}\rightarrow t\bar{t}t\bar{t}$.} \label{fig:2c_mat_smtt}
    \end{subfigure}
    \\
    \begin{subfigure}{0.31\textwidth}
      \includegraphics[width=\linewidth]{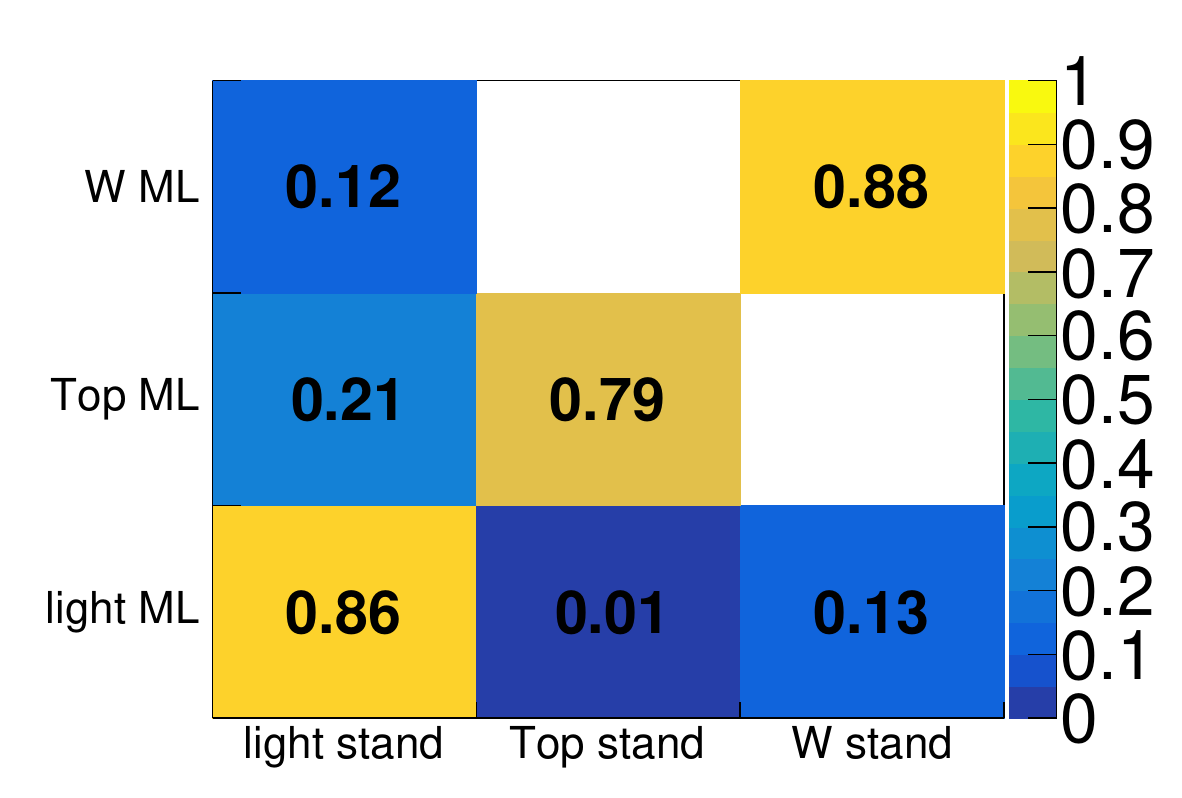}
      \caption{SM $t\bar{t}$.} \label{fig:3a_mat_smtt}
    \end{subfigure}%
    \hspace*{\fill}   
    \begin{subfigure}{0.31\textwidth}
      \includegraphics[width=\linewidth]{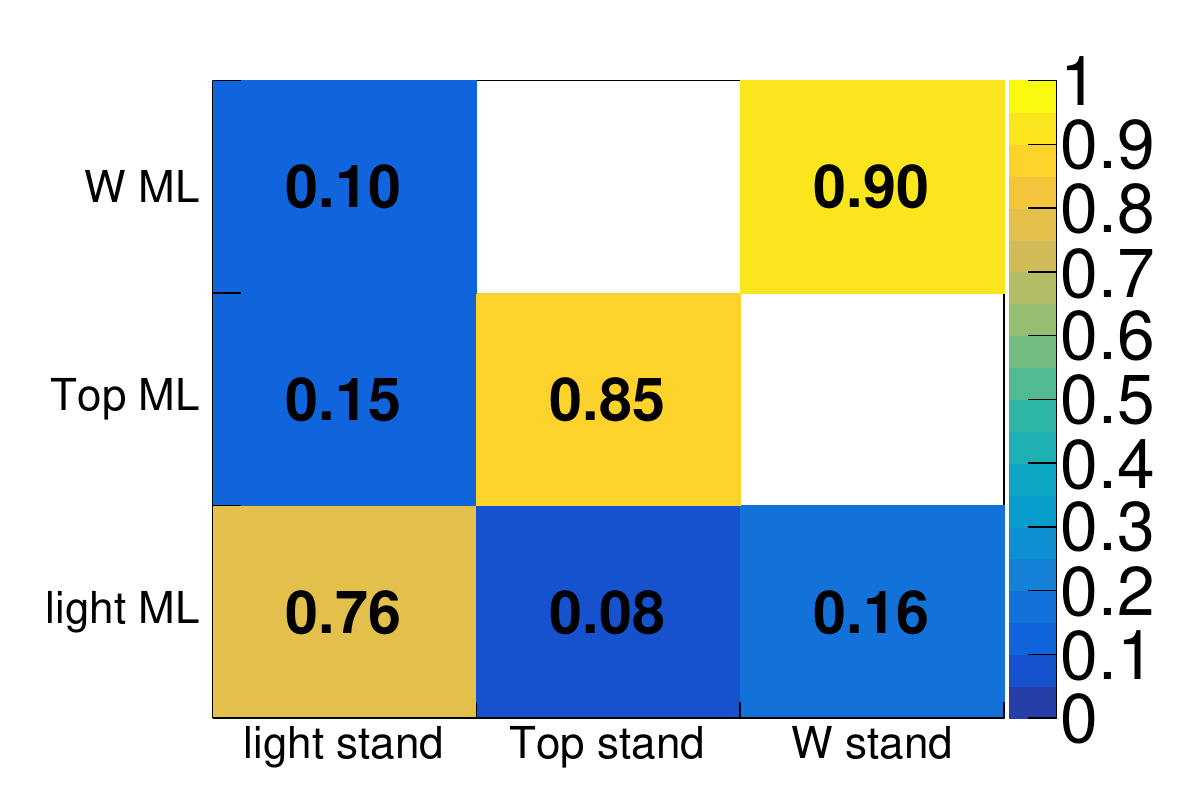}
      \caption{SM $t\bar{t}t\bar{t}$.} \label{fig:3b_mat_smtttt}
    \end{subfigure}%
    \hspace*{\fill}   
    \begin{subfigure}{0.31\textwidth}
      \includegraphics[width=\linewidth]{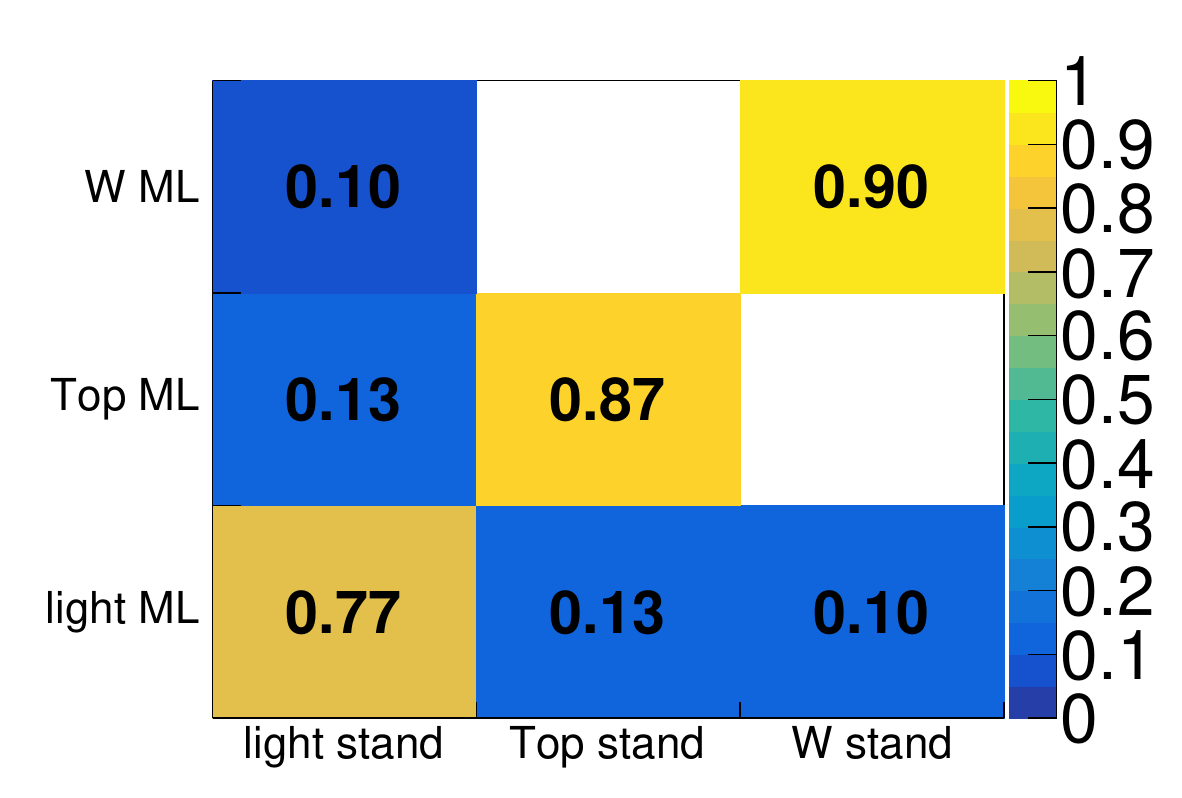}
      \caption{$t\bar{t}y_{0}\rightarrow t\bar{t}t\bar{t}$.} \label{fig:3c_mat_y0}
    \end{subfigure}

  \caption{Confusion matricies of SM $t\bar{t}$ (a, d, g), SM $t\bar{t}t\bar{t}$ (b, e, h), and BSM 
  $t\bar{t}y_{0}\rightarrow t\bar{t}t\bar{t}$ (c, f, i) for cut-based method (a, b, c), ML-based method (d, e, f), 
  and cut-based versus ML-based method (g, h, i). Each matrix is normalized by rows.}
  \label{fig:mat}
\end{figure}

\end{document}